\documentclass[longauth]{aa}
\usepackage{graphicx}
\usepackage[varg]{txfonts}
\usepackage{lscape,longtable}
\usepackage{xcolor}
\usepackage{natbib}

\begin{document}

\title{The Gaia-ESO Survey: Lithium measurements \\
and new curves of growth%
\thanks{Based on observations collected with ESO telescopes at the La Silla
Paranal Observatory in Chile, for the Gaia-ESO Large Public Spectroscopic
Survey (188.B-3002, 193.B-0936, 197.B-1074).}%
\fnmsep\thanks{Tables~A.1--A.3 are fully available in electronic form at the
CDS via anonymous ftp to cdsarc.u-strasbg.fr (130.79.128.5) or via
http://cdsweb.u-strasbg.fr/cgi-bin/qcat?J/A+A/}%
}

\author{%
E. Franciosini\inst{\ref{oaa}} 
\and S. Randich\inst{\ref{oaa}}
\and P. de~Laverny\inst{\ref{nice}}
\and K. Biazzo\inst{\ref{oarm}}
\and D.~K. Feuillet\inst{\ref{lund},\ref{mpifa}}
\and A. Frasca\inst{\ref{oact}}
\and K. Lind\inst{\ref{stockholm}} 
\and L. Prisinzano\inst{\ref{oapa}}
\and G. Tautvai{\v s}ien{\.e}\inst{\ref{vilnius}}
\and A.~C. Lanzafame\inst{\ref{unict}}
\and R. Smiljanic\inst{\ref{ncac}}
\and A. Gonneau\inst{\ref{ioa}}
\and L. Magrini\inst{\ref{oaa}}
\and E. Pancino\inst{\ref{oaa},\ref{ssdc}} 
\and G. Guiglion\inst{\ref{leibniz}}
\and G.~G. Sacco\inst{\ref{oaa}} 
\and N. Sanna\inst{\ref{oaa}}
\and G. Gilmore\inst{\ref{ioa}} 
\and P. Bonifacio\inst{\ref{gepi-m}} 
\and R.~D. Jeffries\inst{\ref{keele}}         
\and G. Micela\inst{\ref{oapa}} 
\and T. Prusti\inst{\ref{estec}}
\and E.~J. Alfaro\inst{\ref{iaa}}        
\and T. Bensby\inst{\ref{lund}} 
\and A. Bragaglia\inst{\ref{oabo}} 
\and P. Fran{\c c}ois\inst{\ref{gepi-p},\ref{upjv}} 
\and A.~J. Korn\inst{\ref{uppsala-oa}} 
\and S. Van~Eck\inst{\ref{ulb}} 
\and A. Bayo\inst{\ref{eso}}
\and M. Bergemann\inst{\ref{mpifa},\ref{nbia}}
\and G. Carraro\inst{\ref{unipd}} 
\and U. Heiter\inst{\ref{uppsala-oa}} 
\and A. Hourihane\inst{\ref{ioa}}            
\and P. Jofr\'e\inst{\ref{portales}} 
\and J. Lewis$^\dagger$\inst{\ref{ioa}} 
\and C. Martayan\inst{\ref{esochile}} 
\and L. Monaco\inst{\ref{unibello}} 
\and L. Morbidelli\inst{\ref{oaa}} 
\and C.~C. Worley\inst{\ref{ioa}} 
\and S. Zaggia\inst{\ref{oapd}}
}

\institute{
INAF -- Osservatorio Astrofisico di Arcetri, Largo E. Fermi 5, 50125
Firenze, Italy \email{elena.franciosini@inaf.it} \label{oaa}
\and
Universit{\'e} C{\^o}te d'Azur, Observatoire de la C{\^o}te d'Azur, CNRS,
Laboratoire Lagrange, Bd de l'Observatoire, CS 34229, 06304 Nice cedex 4,
France \label{nice}
\and 
INAF -- Osservatorio Astronomico di Roma, via Frascati 33, 00040 Monte
Porzio Catone (RM), Italy \label{oarm}
\and
Lund Observatory, Department of Astronomy and Theoretical Physics, Box 43,
SE-22100 Lund, Sweden \label{lund}
\and
Max-Planck Institut f{\"u}r Astronomie, K{\"o}nigstuhl 17, D-69117
Heidelberg, Germany \label{mpifa}
\and
INAF -- Osservatorio Astrofisico di Catania, via S. Sofia 78, 95123 Catania,
Italy \label{oact}
\and
Department of Astronomy, Stockholm University, AlbaNova University Center,
SE-10691 Stockholm, Sweden \label{stockholm}
\and
INAF -- Osservatorio Astronomico di Palermo, Piazza del Parlamento 1, 90134
Palermo, Italy \label{oapa}
\and 
Institute of Theoretical Physics and Astronomy, Vilnius University,
Sauletekio av. 3, LT-10257 Vilnius, Lithuania\label{vilnius}
\and
Dipartimento di Fisica e Astronomia, Sezione Astrofisica, Universit\`a di
Catania, via S. Sofia 78, 95123, Catania, Italy \label{unict}
\and
Nicolaus Copernicus Astronomical Center, Polish Academy of Sciences, ul.
Bartycka 18, 00-716, Warsaw, Poland \label{ncac}
\and
Institute of Astronomy, University of Cambridge, Madingley Road, Cambridge
CB3 0HA, United Kingdom \label{ioa}
\and
Space Science Data Center - Agenzia Spaziale Italiana, via del Politecnico,
s.n.c., 00133, Roma, Italy \label{ssdc}
\and
Leibniz-Institut f{\"u}r Astrophysik Potsdam (AIP), An der Sternwarte 16,
14482 Potsdam \label{leibniz}
\and 
GEPI, Observatoire de Paris, Universit\'e PSL, CNRS, 5 Place Jules Janssen,
F-92195 Meudon, France \label{gepi-m}
\and 
Astrophysics Group, Keele University, Keele, Staffordshire ST5 5BG, United
Kingdom \label{keele}
\and
European Space Agency (ESA), European Space Research and Technology Centre
(ESTEC), Keplerlaan 1, 2201 AZ Noordwijk, The Netherlands \label{estec}
\and
Instituto de Astrof\'{i}sica de Andaluc\'{i}a, CSIC, Glorieta de la
Astronom\'{i}a SNR, 18008 Granada, Spain \label{iaa}
\and
INAF -- Osservatorio di Astrofisica e Scienza dello Spazio, via P. Gobetti
93/3, 40129 Bologna, Italy \label{oabo}
\and
GEPI, Observatoire de Paris, PSL Research University, CNRS, 61 avenue de
l'Observatoire, 75014 Paris, France \label{gepi-p}
\and
UPJV, Universit\'e de Picardie Jules Verne, 33 rue St Leu, 80080 Amiens,
France \label{upjv}
\and
Observational Astrophysics, Division of Astronomy and Space Physics,
Department of Physics and Astronomy, Uppsala University, Box 516, SE-75120
Uppsala, Sweden \label{uppsala-oa}
\and
Institut d'Astronomie et d'Astrophysique, Universit\'{e} libre de Bruxelles,
CP 226, Boulevard du Triomphe, 1050 Bruxelles, Belgium \label{ulb} 
\and
European Southern Observatory, Karl-Schwarzschild-Strasse 2, 85748 Garching
bei M{\"u}nchen, Germany \label{eso}
\and
Niels Bohr International Academy, Niels Bohr Institute, University of
Copenhagen, Blegdamsvej 17, DK-2100 Copenhagen {\O}, Denmark \label{nbia}
\and
Dipartimento di Fisica e Astronomia, Universit\`a di Padova, Vicolo
dell'Osservatorio 2, 35122 Padova, Italy \label{unipd}
\and
N\'ucleo de Astronom\'{i}a, Facultad de Ingenier\'{i}a y Ciencias,
Universidad Diego Portales, Av. Ej\'ercito 441, Santiago, Chile
\label{portales}
\and 
European Organisation for Astronomical Research in the Southern Hemisphere,
Alonso de C{\'o}rdova 3107, Vitacura,  Casilla 19001, Santiago 19, Chile
\label{esochile}
\and
Departamento de Ciencias Fisicas, Universidad Andres Bello, Fernandez
Concha 700, Las Condes, Santiago, Chile \label{unibello}
\and
INAF -- Osservatorio Astronomico di Padova, Vicolo dell'Osservatorio 5,
35122 Padova, Italy \label{oapd}
}

\date{Received ... / Accepted ...}

\abstract%
{The Gaia-ESO Survey (GES) is a large public spectroscopic survey that was
carried out using the multi-object FLAMES spectrograph at the Very Large
Telescope. The survey provides accurate radial velocities, stellar
parameters, and elemental abundances for $\sim$\,115\,000 stars in all Milky
Way components.}
{In this paper we describe the method adopted in the final data release to
derive lithium equivalent widths (EWs) and abundances.}
{Lithium EWs were measured using two different approaches for FGK and M-type
stars, to account for the intrinsic differences in the spectra. For FGK
stars, we fitted the lithium line using Gaussian components, while direct
integration over a predefined interval was adopted for M-type stars. Care
was taken to ensure continuity between the two regimes. Abundances were
derived using a new set of homogeneous curves of growth that were derived
specifically for GES, and which were measured on a synthetic spectral grid
consistently with the way the EWs were measured. The derived abundances were
validated by comparison with those measured by other analysis groups using
different methods.}
{Lithium EWs were measured for $\sim$\,40\,000 stars, and abundances could
be derived for $\sim$\,38\,000 of them. The vast majority of the measures
(80\%) have been obtained for stars in open cluster fields. The remaining
objects are stars in globular clusters, or field stars in the Milky Way
disc, bulge, and halo.}
{The GES dataset of homogeneous lithium abundances described here will be
valuable for our understanding of several processes, from stellar evolution
and internal mixing in stars at different evolutionary stages to Galactic
evolution.}


\titlerunning{GES lithium measurements and new curves of growth}
\maketitle


\section{Introduction}
\label{sec:intro}

Lithium is a key element for our understanding of several open issues in
astrophysics, from Big-Bang nucleosynthesis to the chemical evolution of the
Milky Way, to mixing processes in stellar interiors and stellar evolution
\citep[e.g.][and references therein]{rm21}. Precise and homogeneous measures
for large samples of stars are fundamental to address these issues.

The Gaia-ESO Survey \citep[hereafter GES;][]{GES12,GES13} is a large public
spectroscopic survey that observed $\sim$\,$10^5$ stars in all Milky Way
components, from the thin and thick disc to the bulge and halo, including 65
(science and calibration) open clusters and 15 globular clusters. The GES
spectra were acquired with the Fiber Large Array Multi-Element Spectrograph
\citep[FLAMES;][]{pasquini02} mounted on the UT2 unit of the Very Large
Telescope, using the high-resolution Ultraviolet and Visual Echelle
Spectrograph (UVES) and the Giraffe instrument operated in MEDUSA mode. The
dataset was complemented with additional spectra of stars in open and
globular clusters retrieved from the ESO archive and observed with the same
setups, increasing the total number of open clusters to 83. The final
dataset provides precise radial velocities and homogeneous stellar
parameters and chemical abundances of up to 31 elements, including lithium,
for $\sim$\,115\,000 stars. 

A detailed overview of the survey goals and strategy, and of the data
analysis is provided by \citet{gilmore22} and \citet{randich22}. More
specific papers describe the target selection
\citep{stonkute16,bragaglia22}, the calibration strategy \citep{pancino17},
the data reduction pipelines, and the derivation of radial and rotational
velocities \citep{sacco14,jeffries14,jackson15}. Spectral analysis of
late-type stars was performed by different analysis nodes within each of the
three dedicated working groups: WG10 and WG11 for FGK stars observed with
Giraffe and UVES, respectively, and WG12 for pre-main-sequence (PMS) stars
\citep[Worley et al. in prep.]{smiljanic14,lanzafame15}. The individual
results were then homogenised and combined to produce the final recommended
values. The analysis was performed in two steps: first, atmospheric
parameters were derived and homogenised, then chemical abundances were
obtained using the recommended atmospheric parameters as input.

In this paper we describe the derivation of lithium abundances for the Sixth
internal Data Release (iDR6), which corresponds to the final data release
published in the ESO archive%
\footnote{\url{https://www.eso.org/qi/catalogQuery/index/393}.}.
In iDR6, recommended lithium abundances were only derived by the Arcetri
node. This approach differs from previous data releases, where a multi-node
analysis was performed just as for the other parameters and abundances
\citep[see][]{smiljanic14,lanzafame15}. Because of the peculiarity of
lithium, which is measured by a single line and depends on stellar mass and
age, and of the different subsets of spectra that the individual nodes were
able to analyse, the homogenisation of the results obtained from different
nodes did not prove to be efficient in providing sufficiently homogeneous
results. However, a sub-sample of the spectra was still analysed by other
nodes, and these measurements were used for validation purposes.

The method adopted by the Arcetri node is based on the measurement of
equivalent widths (EWs) and the use of curves of growth (COGs) to derive the
abundances. The advantage of the EW method over spectral synthesis is that
EWs can also be measured for spectra where spectral synthesis is not
feasible, or when not all atmospheric parameters can be derived (e.g. for
low signal-to-noise spectra, or young rapidly rotating stars). However, to
ensure that no bias is introduced, it is preferable that both EWs and COGs
are derived in a consistent way. For this reason, a new set of COGs,
specific for GES, was also derived.

The paper is organised as follows. In Sect.~\ref{sec:data} we present the
lithium data available in GES. The derivation of the new set of COGs and the
lithium measurements are described in Sects.~\ref{sec:cogs} and
\ref{sec:measures}, respectively. In Sect.~\ref{sec:validation} we discuss
the validation of the results. The final catalogue is presented in
Sect.~\ref{sec:catalogue}, and caveats for the measures are given in
Sect.~\ref{sec:caveats}. A summary is provided in Sect.~\ref{sec:summary}.


\section{Lithium in GES}
\label{sec:data}

Lithium can be measured from the \ion{Li}{i} doublet at 6707.8~\AA{}, which
is available in GES spectra acquired with the UVES U580 setup (480--680~nm,
$R=47\,000$) and with the Giraffe HR15N grating (644--680~nm,
$R\sim$\,17\,000). The UVES U580 setup was used for FGK turn-off and giant
stars in old open clusters, globular clusters, and the Milky Way fields, and
for the brightest main-sequence stars in young open clusters. Giraffe HR15N
spectra were acquired for FGK stars on the main sequence in open and
globular clusters, and for G- to M-type PMS stars in young open clusters. A
fraction of the targets was observed with both instruments for
cross-calibration purposes \citep[see][]{bragaglia22}.

\begin{figure}
\centering
\resizebox{\hsize}{!}{\includegraphics{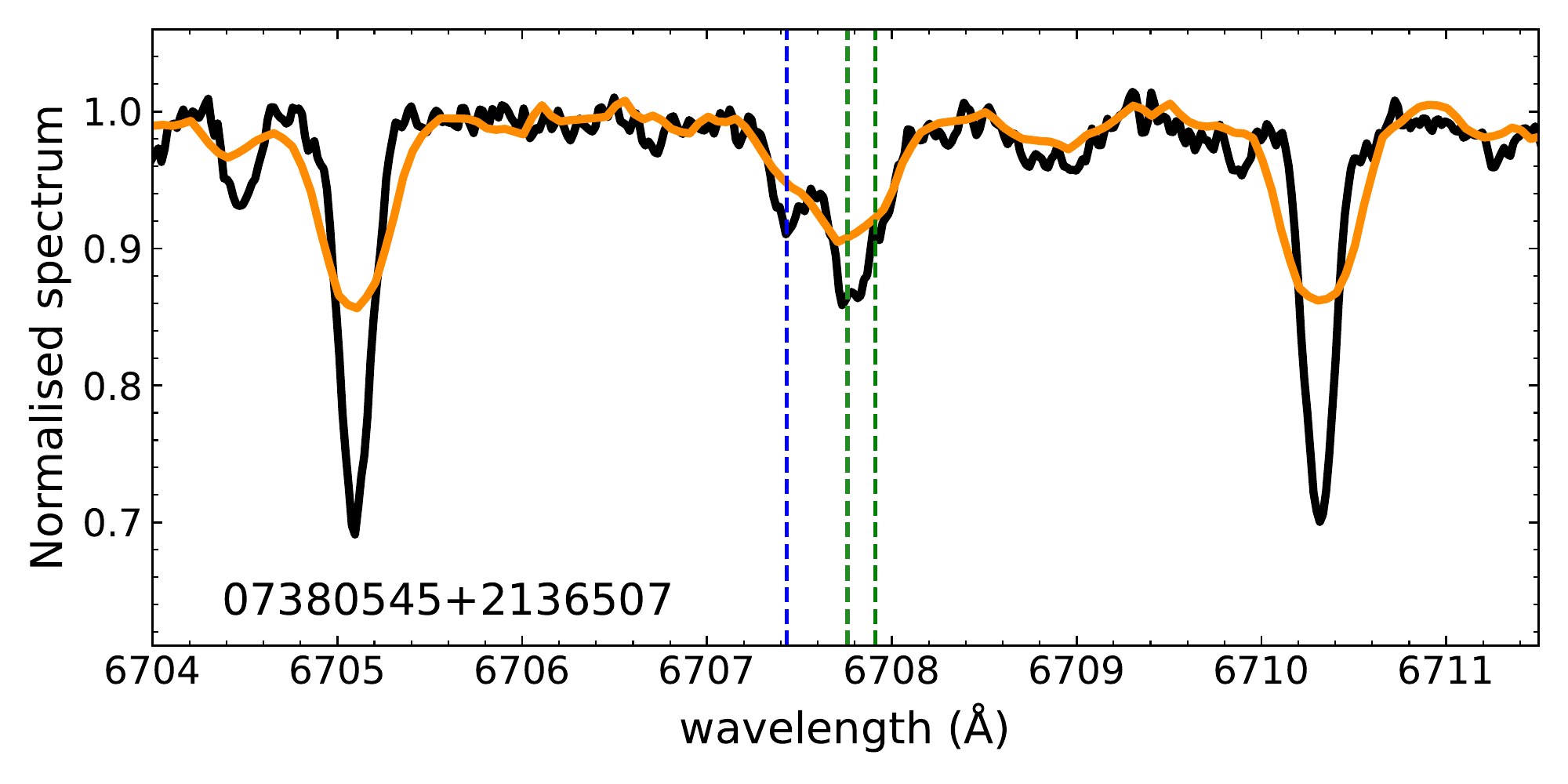}}
\caption{UVES (black, S/N=119) and Giraffe (orange, S/N=223) spectra of the
giant star 07380545+2136507 which was observed with both instruments. The
vertical green and blue dashed lines mark the positions of the lithium
doublet and of the 6707.4~\AA{} iron line, respectively.}
\label{fig:compare_ug}
\end{figure}

\begin{figure}
\centering
\resizebox{\hsize}{!}{\includegraphics{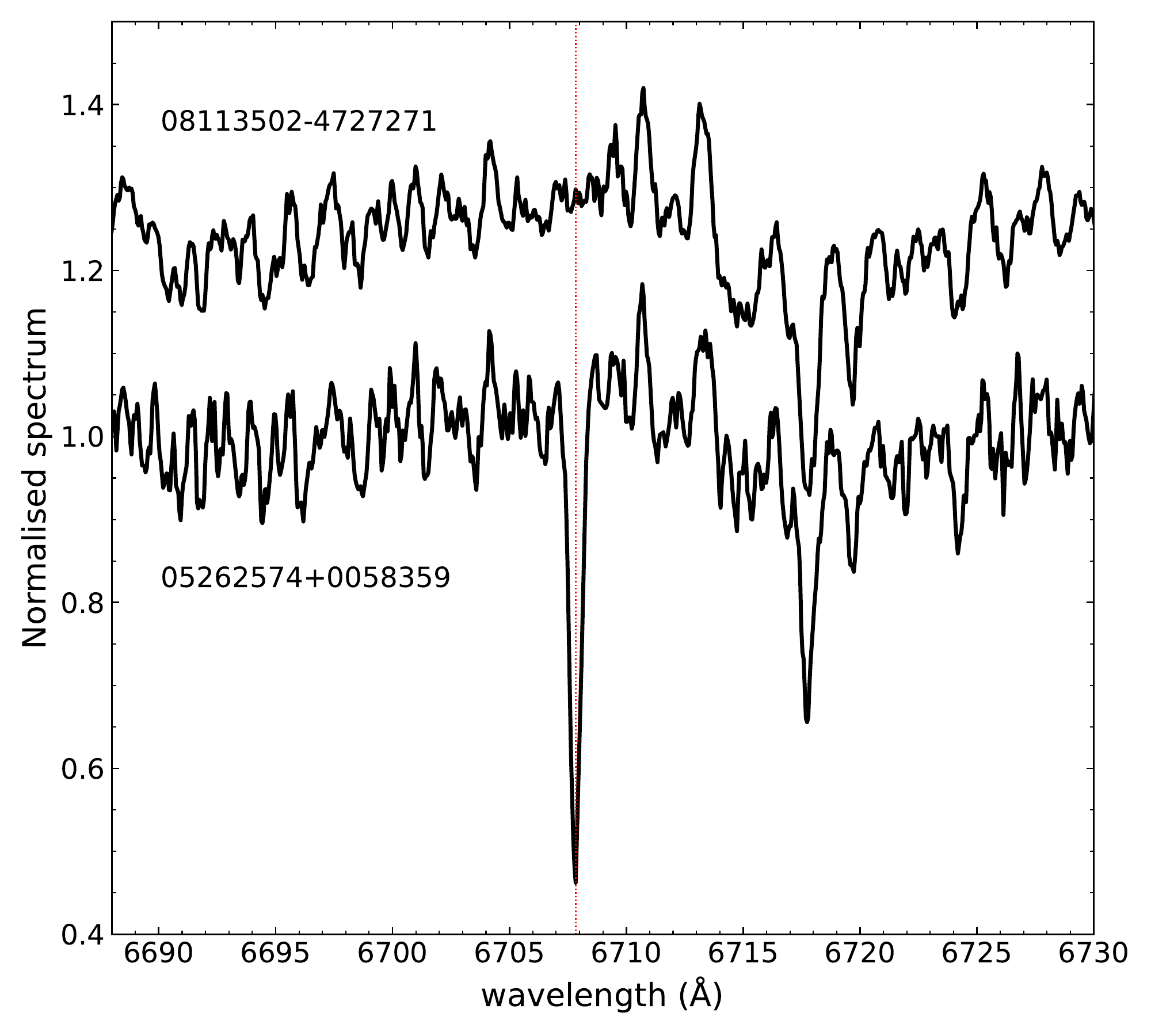}}
\caption{Spectra of two M-type dwarf stars with $T_\mathrm{eff}\sim 3500$~K
and solar metallicity observed with Giraffe. The upper and lower spectrum
correspond to a fully depleted star ($A(\mathrm{Li})<-1.0$) and to a star
with $A$(Li)\,$\sim$\,2.0, respectively. The spectra are normalised to their
average flux and the upper one has been arbitrarily shifted upwards by 0.25
for clarity. The vertical red dotted line marks the mean position of the
lithium line.}
\label{fig:spectra_M}
\end{figure}

The large variety of the stellar samples targeted by GES makes a homogeneous
determination of lithium abundances challenging. In the case of FGK stars,
the lithium line is blended mainly with the nearby \ion{Fe}{i} line at
6707.43~\AA, plus a few weaker components from other elements. At the UVES
resolution, it is generally possible to deblend the lines and measure
directly the lithium-only EW, EW(Li). Exceptions to this are stars with high
rotation rates ($v\sin i \ga 30$~km~s$^{-1}$), or young stars with very
strong lithium, where the deblending might not be possible. The deblending
is never possible for Giraffe HR15N spectra. In such cases, only the total
EW(Li+Fe) can be measured. An example of this is shown in
Fig.~\ref{fig:compare_ug}, where we compare the spectra of a giant star
observed with both instruments: while in UVES the iron and lithium lines,
although partially overlapping, are clearly distinct, this is not the case
for Giraffe.

In M-type stars, the measurement of lithium abundances is complicated by the
presence of molecular bands and additional lines from other elements, which
are severely blended with lithium and cause a strongly depressed
pseudo-continuum \citep[e.g.][]{zapatero02}. The pseudo-continuum trend can
be clearly seen in Fig.~\ref{fig:spectra_M}, where we compare the spectrum
of an M-type star with lithium abundance\footnote{%
In the usual notation $A$(Li)$\,=\log N(\mathrm{Li})/N(\mathrm{H})+12$} 
$A$(Li)\,$\sim$\,2.0, with that of a fully depleted star with similar
temperature and rotation; the wide depression around the lithium position
caused by the other line blends is clearly evident in the latter. This
pseudo-continuum masks the position of the real continuum level, preventing
the measure of the true lithium EW, contrary to what can be done in hotter
stars. In this case, only a pseudo-EW (pEW) can be derived. 

The above considerations imply that different approaches must be adopted to
measure lithium in FGK- and M-type stars. However, the chosen method must
ensure the highest possible consistency between the measures of the two sets
of stars, minimising any discontinuity between the two regimes. To this aim,
we developed a specific code based on python to measure EWs and pEWs and the
corresponding COGs in a consistent way.


\section{New lithium curves of growth}
\label{sec:cogs}

The need for a new set of homogeneous COGs for lithium arises from the fact
that COGs covering the entire range of atmospheric parameters of GES spectra
are not available in the literature. As mentioned by \citet{lanzafame15}, in
previous data releases we used the COGs derived by \citet{soderblom93} for
$T_\mathrm{eff} = 4000-6500$~K, and those by \citet{palla07} for
$T_\mathrm{eff} \le 4000$~K. However, these COGs are only valid for dwarfs,
therefore they are not applicable to giants; they were computed for solar
metallicity and are therefore not valid for more metal-poor or metal-rich
stars; and they were derived using inconsistent methods, causing
discontinuity problems around 4000~K. Moreover, literature COGs were derived
using different sets of model atmospheres, which are not necessarily the
same used in GES. 

To derive the new set of COGs, we used a grid of synthetic spectra that was
computed as in \citet{delaverny12} and \citet{guiglion16}, within the
context of the AMBRE project \citep{delaverny13}. The grid is based on the
same model atmospheres used in GES, namely the one-dimensional MARCS models
in local thermodynamic equilibrium \citep[LTE;][]{gustafsson08}, and was
computed assuming standard [$\alpha$/Fe] ratios\footnote{%
[$\alpha$/Fe]\,$=0.0$ for $\mathrm{[M/H]}\ge 0.0$, [$\alpha$/Fe]\,$=+0.4$
for $\mathrm{[M/H]}\le -1.0$, and following a linear trend in-between.}. 
The parameters were chosen to cover the entire range of late-type stars
observed in GES, namely $3000 \le T_\mathrm{eff} \le 8000$~K, $0.5 \le \log
g \le 5.0$ and $-2.50 \le \mathrm{[Fe/H]} \le +0.50$. Lithium abundances
vary from $A(\mathrm{Li}) = -1.0$ to $A(\mathrm{Li}) = +4.0$ in steps of
0.2~dex, except for $\mathrm{[Fe/H]} < -1.50$, where abundances have been
limited to $A(\mathrm{Li}) \le +3.4$, since higher abundances at such low
metallicities are extremely rare \citep[e.g.][]{sanna20}.

As mentioned in the previous section, the differences in the spectra of FGK
and M-type stars require the adoption of two different approaches to measure
the EWs. For this reason, two separate sets of COGs were derived, one for
FGK stars, covering the temperature range 4000~K$\,\le T_\mathrm{eff}\le
8000$~K, and a separate set for M-type stars, covering the range
3000~K$\,\le T_\mathrm{eff}\le 4500$~K. The two sets overlap over the range
$4000-4500$~K to ensure continuity. The M-type COGs were however limited to
[Fe/H]\,$\ge -1.5$, since lower-metallicity M-type stars are generally not
present in GES.

\subsection{FGK stars}
\label{sec:cogs_FGK}

In the case of FGK stars, which were observed with both UVES and Giraffe, we
need a set of COGs that can equally be applied to cases where we can
directly measure the deblended lithium line and to those where only the
blended line can be measured. To achieve this, COGs were derived for the
lithium component only, and a corresponding set of corrections for the Fe
blend was computed. These corrections can be applied to derive the Li-only
EW when only the total EW(Li+Fe) can be measured, before computing the
abundances.

The EWs of the lithium doublet and of the Fe blend were measured on the
spectral grid degraded to the UVES resolution, down to
$T_\mathrm{eff}=4000$~K. For simplicity, and to ease the deblending of the
two components, for each set of atmospheric parameters both Li and Fe EWs
were simultaneously measured only on the spectrum with $A(\mathrm{Li}) =
-1.0$, which was used as reference spectrum. The lines were fitted with
Gaussian components, assuming a local linear continuum, using the code
described in Sect.~\ref{sec:ews}. The EWs of the Fe blend obtained from this
fit were taken as the blend correction for the corresponding parameters. For
higher lithium abundances, the lithium EW was measured by spectral
subtraction, that is, we subtracted the reference spectrum from the
corresponding spectra with $A(\mathrm{Li})>-1.0$, and integrated the
resulting residual components. We chose to integrate the residuals, instead
of fitting them, because at the highest lithium abundances the adopted
Gaussian components are not able to fit the residual line correctly,
especially below 4500~K, where the lithium line develops significant wings
and is better described by a Voigt profile. The half-width of the
integration interval was set to $d=3.5\sigma$, where $\sigma$ is the width
of the corresponding best-fit Gaussian, with a maximum allowed value of
0.8~\AA. The latter value corresponds to the typical width of the core of
the line for $A(\mathrm{Li})>3.0$, and is also consistent with the typical
full width of the line in Giraffe spectra (see Sect.~\ref{sec:cogs_M}). The
resulting EW was summed to the corresponding value measured for
$A(\mathrm{Li}) = -1.0$ to derive the final lithium EW. The COGs and the
blend corrections are given in Tables~\ref{tab:cog_fgk} and
\ref{tab:fecorr}, respectively. 

\begin{figure}
\centering
\resizebox{\hsize}{!}{\includegraphics[clip]{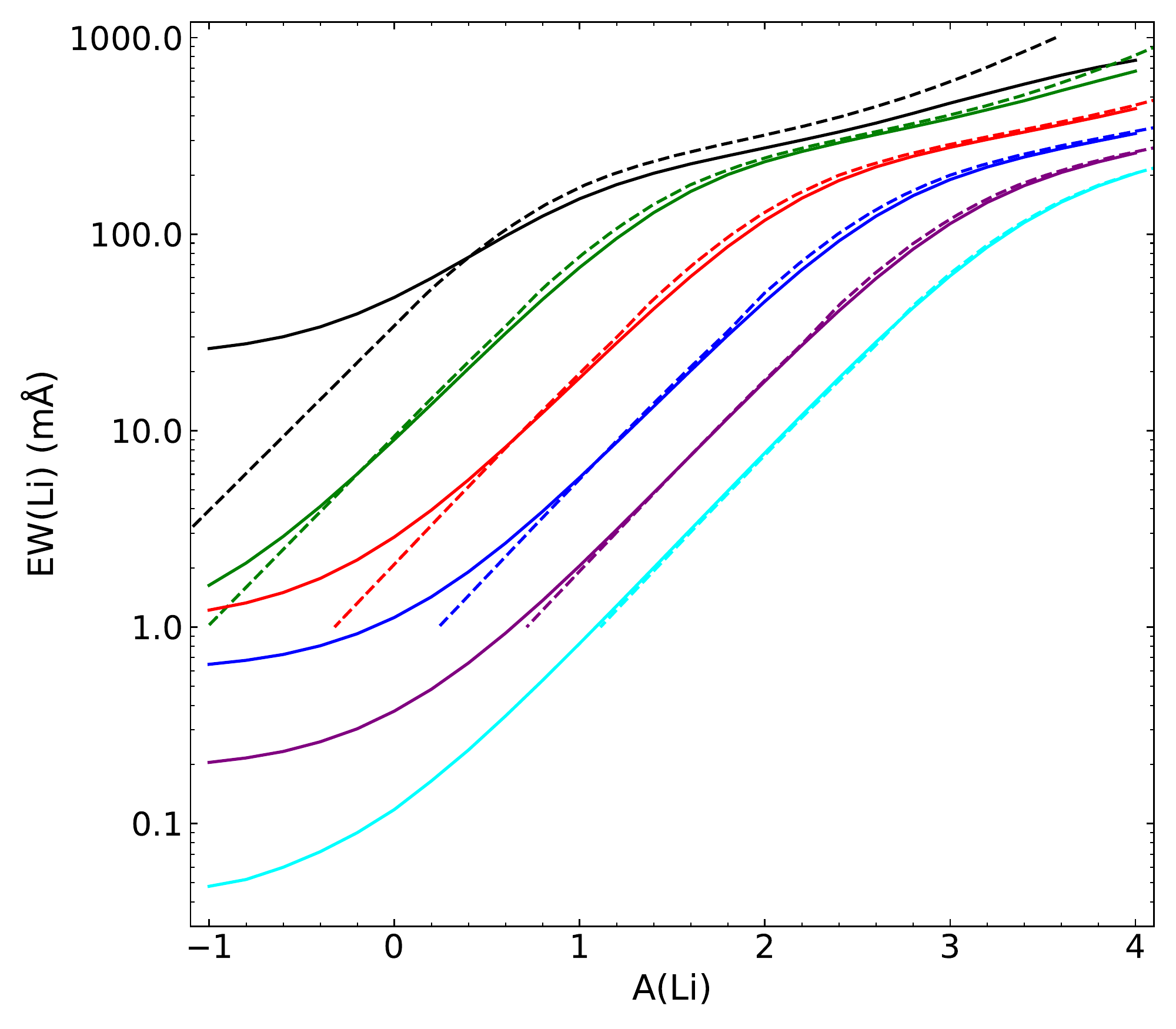}}
\caption{Comparison of the COGs from \citet[dashed lines]{soderblom93} with
our FGK COGs (solid lines) for the corresponding parameters ($\log g=4.5$
and solar metallicity). COGs are plotted for $T_\mathrm{eff}=4000$, 4500,
5000, 5500, 6000, and 6500~K (from top to bottom).}
\label{fig:sod}
\end{figure}

In Fig.~\ref{fig:sod} we compare the COGs for solar metallicity and $\log g
= 4.5$ with those of \citet{soderblom93}, for different values of
$T_\mathrm{eff}$ between 4000 and 6500~K. There is an excellent agreement
between the two sets, except at low abundances or for $T_\mathrm{eff} =
4000$~K, likely due to the differences in the way the COGs were measured.
The discrepancies at 4000~K are also due to the difficulty of deblending the
individual Li and Fe lines from the other line and molecular blends at solar
metallicity.

\subsection{M-type stars}
\label{sec:cogs_M}

According to the Milky Way and cluster target selection function
\citep[see][]{stonkute16,bragaglia22}, M-type stars in GES are generally
observed only with Giraffe HR15N and not with UVES (except for a handful of
objects), and they are mainly young objects that may have large rotation
rates. Therefore COGs in this regime are required for Giraffe HR15N only,
and rotation must also be taken into account.

\begin{figure}
\centering
\resizebox{\hsize}{!}{\includegraphics{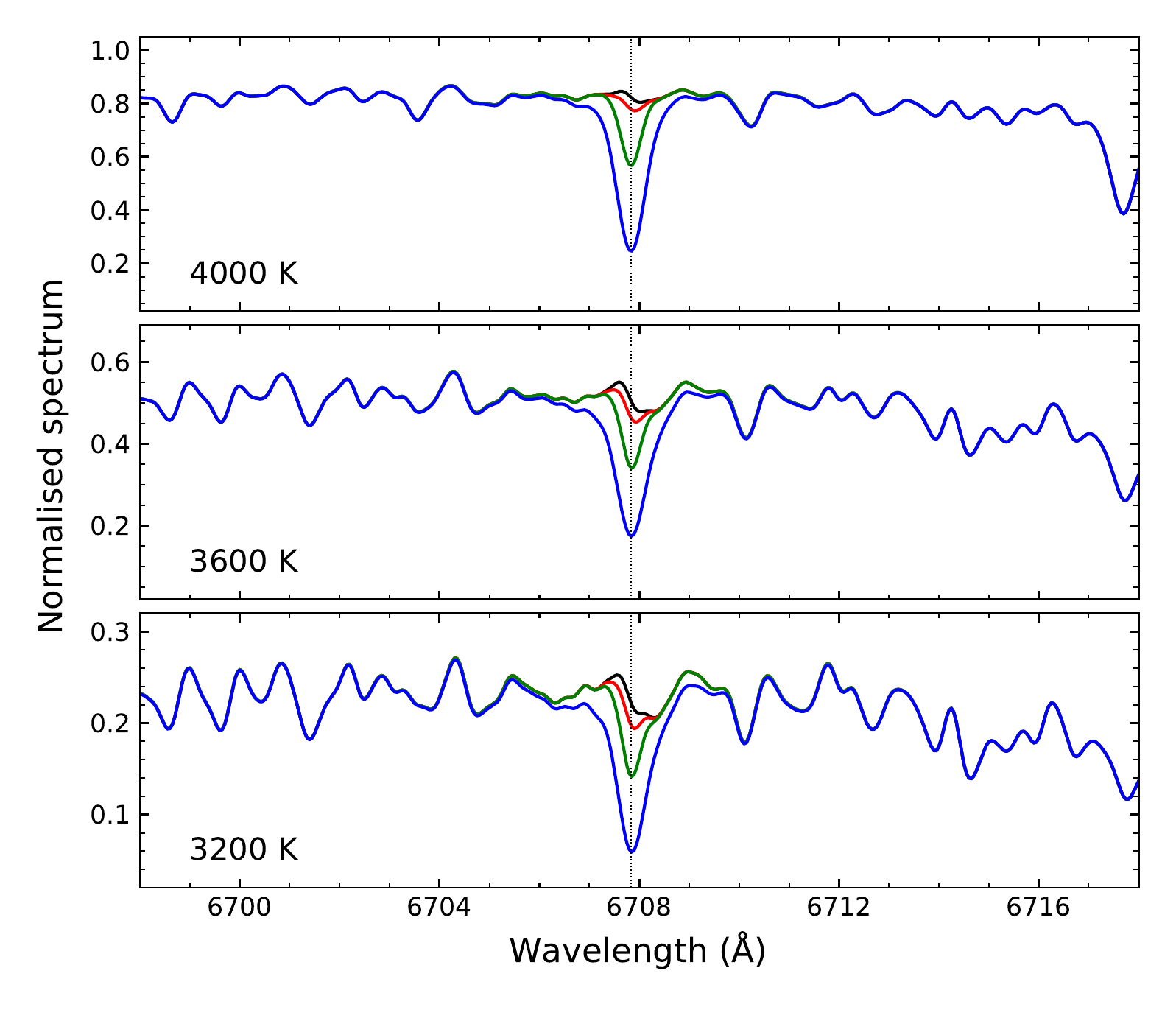}}
\caption{Synthetic spectra of M-type dwarf stars with $T_\mathrm{eff}=4000$,
3600 and 3200~K (from top to bottom), $\log g=4.5$ and solar metallicity, at
the Giraffe resolution. Spectra are plotted for $A\mathrm{(Li)}=-1.0$ (black
line), 0.0 (red line), $+1.0$ (green line), and $+3.0$ (blue line). The
vertical dotted line shows the mean position of the lithium doublet.}
\label{fig:mtype_synt}
\end{figure}

To measure the COGs, we degraded the spectral grid for $T_\mathrm{eff} \le
4500$~K at the Giraffe resolution. In Fig.~\ref{fig:mtype_synt} we show an
example of the synthetic spectra for three temperatures and different
lithium abundances. The figure clearly shows the increasing depression of
the pseudo-continuum level going towards lower temperatures, and the strong
blending with other components that dominate the absorption when the lithium
line is weak or not present. A local pseudo-continuum was defined as the
envelope of the molecular bands, approximated by a straight line passing
through the maxima of the spectra on both sides of the lithium line.
Analysis of the synthetic spectra shows that these maxima are generally
located in the wavelength intervals [6703.0,\,6705.0]~\AA{} and
[6710.0,\,6712.0]~\AA{} for slow rotators. To account for the different
rotation rates of the target stars, COGs were measured for nine different
rotational velocities from 0 to 150~km~s$^{-1}$. However, the full set of
rotational velocities could only be measured for $A(\mathrm{Li})> 2.0$. At
lower abundances the maximum $v\sin i$ was set to lower values, as indicated
in Table~\ref{tab:maxrot_M}. We also limited $v\sin i$ to 50~km~s$^{-1}$ for
giants with $\log g<3.5$, since higher rotational velocities are generally
not found in these stars.

\begin{table}
\centering
\caption{Maximum allowed $v\sin i$ in M-type COGs}
\label{tab:maxrot_M}
\begin{tabular}{cc}
\hline\hline\noalign{\smallskip}
Max. $v\sin i$& Range \\
(km~s$^{-1}$)& \\
\noalign{\smallskip}\hline\noalign{\smallskip}
150& $A(\mathrm{Li})> 2.0$\\
100& $1.0< A(\mathrm{Li})< 2.0$\\
 50& $0.0 < A(\mathrm{Li})< 1.0$\\
 30& $A(\mathrm{Li})< 0.0$\\
\noalign{\smallskip} 
50& $\log g < 3.5$\\
\noalign{\smallskip}\hline
\end{tabular}
\end{table}

The pEW was then derived by direct integration within a specified wavelength
interval, which was defined as a symmetric interval of width $\pm d$ around
$\lambda_\mathrm{Li} = 6707.84$~\AA. The value of $d$ was derived by
measuring the full width of the lithium line in the synthetic spectra, and
was also verified using observed spectra. For slow-rotators we found
$d=0.8$~\AA: this width is appropriate up to a projected rotational velocity
$v\sin i = 20$~km~s$^{-1}$. Above this threshold, the line broadens
according to the following relation:
\begin{equation}
d=0.8 + 0.02 \,(v \sin i -20)\; \text{\AA}.
\end{equation}
We checked that this interval is also consistent with the typical width of
the lithium line in Giraffe spectra of FGK stars, hence no major
inconsistencies are expected between the two regimes. The derived COGs are
given in Table~\ref{tab:cog_m}.

\begin{figure*}[!ht]
\resizebox{\hsize}{!}{\includegraphics[clip]{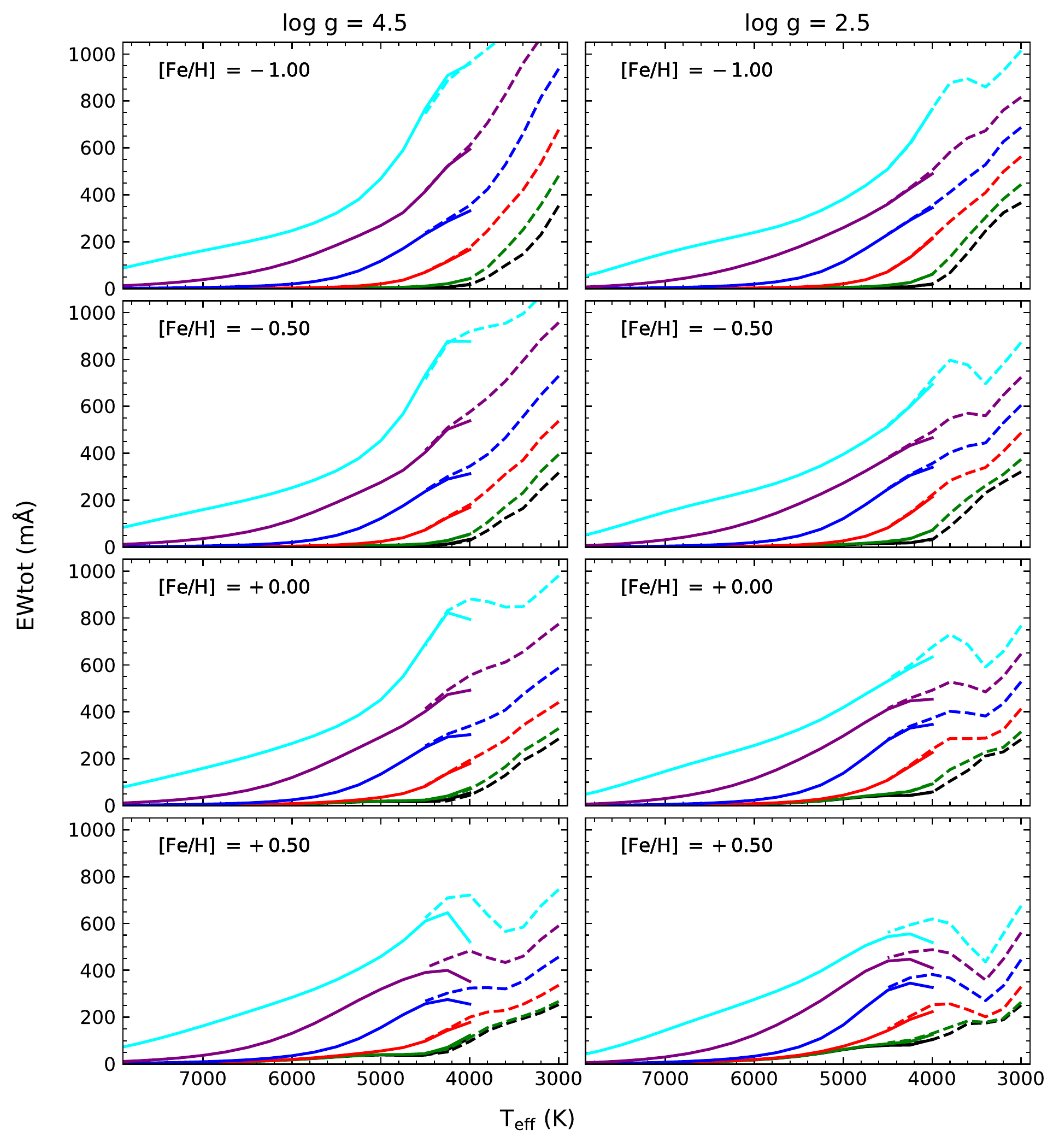}}
\caption{Comparison of the measured total EW (Li+Fe) vs $T_\mathrm{eff}$ for
FGK (solid lines) and M-type stars with $v\sin i=0$ (dashed lines), for
$\log g = 4.5$ (left panels) and 2.5 (right panels), and different
metallicities. The COGs are plotted for $A(\mathrm{Li}) = -1.0$ (black), 0.0
(green), $+$1.0 (red), $+$2.0 (blue), $+$3.0 (purple), and $+$4.0 (cyan).}
\label{fig:cogs}
\end{figure*}

Figure~\ref{fig:cogs} compares the two sets of COGs for dwarfs ($\log
g=4.5$) and giants ($\log g=2.5$) at different metallicities. For the FGK
regime, the total Li+blends EW is plotted. The plots show that the two sets
connect smoothly, implying that homogeneity is ensured. As mentioned before,
the discrepancies seen at 4000-4250~K are due to the difficulty of measuring
the individual Li and Fe EWs at these temperatures, especially at higher
metallicities where molecular blends are already important. This problem
appears to be stronger for dwarfs than for giants, because of the stronger
molecular lines present in dwarf stars. For this reason, in this temperature
interval it is advisable to adopt the integration method and the M-type
COGs, if possible. The dip seen at $\sim$\,3500~K was already noted by
\citet{palla07}, and is likely due to changes in the relative contribution
of TiO bands and lithium. We also note that, because of the presence of the
additional line components blended with lithium that are included in the
integration interval (see Fig.~\ref{fig:mtype_synt}), the pEW can never be
equal to zero, even when no lithium line is present. For a non-rotating
dwarf star at solar metallicity the pEW for $A(\mathrm{Li})=-1.0$ ranges
from $\sim$\,50~m\AA{} at 4000~K to $\sim$\,300~m\AA{} at 3000~K.


\section{Lithium measurements}
\label{sec:measures}

Lithium measurements on GES spectra were performed as consistently as
possible with the way the COGs were derived. The EWs and pEWs were generally
measured only for stars with signal-to-noise ratio S/N$\,>10$ (or 20 in some
cases), except for young stars with strong lithium absorption, where we
lowered the limit to S/N$\,>5$ whenever possible. In addition, measures were
generally limited to $v\sin i \la 100$~km~s$^{-1}$ in Giraffe and
50~km~s$^{-1}$ in UVES, unless the line was sufficiently strong to be also
measurable at higher rotation rates\footnote{%
These thresholds were applied to the rotational velocities derived by the
data reduction pipelines for the individual spectra. For part of the stars
these values have been superseded by more accurate measurements after the
homogenisation process, therefore these conditions may not be strictly met
in the final catalogue. While our code in most cases handled rotation
correctly, we caution that some measures with $v\sin i>150$~km~s$^{-1}$ and
EW(Li+Fe)$\,<100$~m\AA{} could be inaccurate.}. 
The lower threshold adopted for UVES is due to the fact that the bulk of the
stars observed with the U580 setup have $v\sin i < 20$~km~s$^{-1}$, and the
number of objects above the threshold with a measurable Li line is small. We
also excluded spectra with clear evidence of double (or multiple) lines,
indicative of SB2 binaries, or SB1 binaries with large velocity variations,
for which the derived parameters are likely not accurate. Before performing
the measurements, spectra were shifted to a rest frame based on their radial
velocities.

\begin{figure*}[!ht]
\centering
\resizebox{\hsize}{!}{\includegraphics{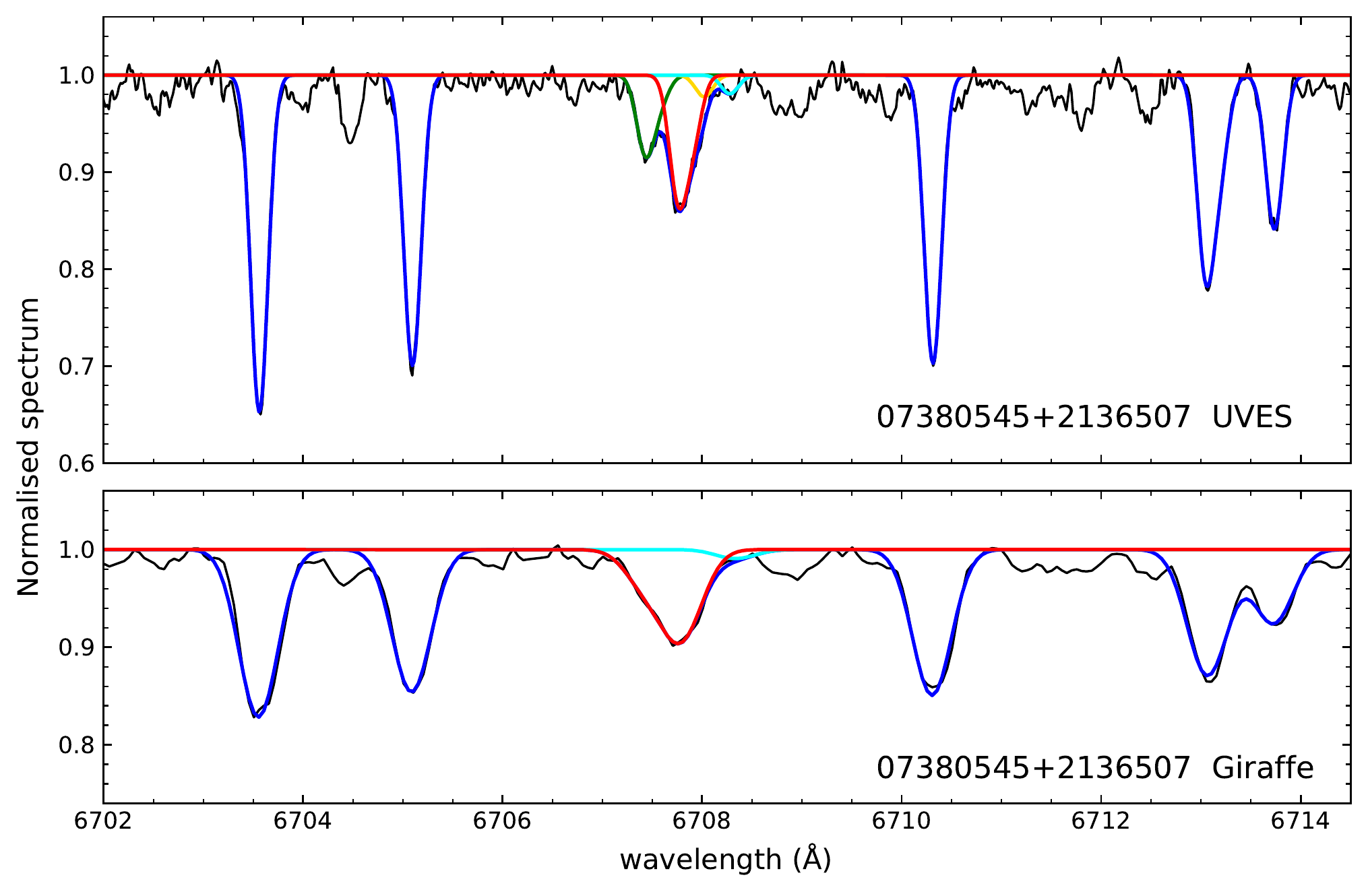}}
\caption{Best-fit results for the UVES (upper panel, S/N\,=\,119) and
Giraffe (lower panel, S/N\,=\,223) spectra of the giant star
07380545+2136507 shown in Fig.~\ref{fig:compare_ug} ($T_\mathrm{eff}=4899\pm
34$~K, $\log g = 2.80\pm 0.05$, $\mathrm{[Fe/H]}=-0.17\pm 0.05$). In
the upper panel, the red and green lines show the best-fit lithium and Fe
blend components, respectively, while in the lower panel only the combined
Li+Fe component is shown in red. The gold and cyan lines are the best-fit
components at $\sim$\,6708.0 and 6708.3~\AA{}. The total fit including the
additional lines (see text) is marked in blue. The measured EWs are
EW(Li)$\,=40.0\pm 0.8$~m\AA, EW(Fe)$\,=22.8\pm 0.8$~m\AA{} for UVES, and
EW(Li+Fe)$\,=64.6\pm 1.1$~m\AA{} for Giraffe. The resulting lithium
abundances are $A(\mathrm{Li})=1.25\pm 0.04$ for UVES and $1.30\pm
0.05$ for Giraffe.}
\label{fig:fit_ug}
\end{figure*}

\subsection{Continuum level}
\label{sec:cont}

As mentioned in Sects.~\ref{sec:cogs_FGK} and \ref{sec:cogs_M}, the local
continuum or pseudo-continuum was approximated as a straight line, passing
through the maxima of the spectrum at the two sides of the lithium line.
These maxima were searched in the intervals [6701.0,\,6705.5]~\AA{} and
[6709.5,\,6715.0]~\AA{} for FGK stars, and in the intervals defined in
Sect.~\ref{sec:cogs_M} for M-type stars. The larger intervals used for FGK
stars account for the blending of lines in Giraffe spectra, which can
slightly depress the continuum around lithium in high-metallicity or giant
stars. To identify the position of the maxima, we performed a non-parametric
fit of the spectra within the two intervals, using a median
filter\footnote{To this aim, we used the {\tt scipy.signal.medfilt} code.} 
with a variable smoothing window depending on the S/N, and calibrated
separately for UVES and Giraffe (to account for the different resolution),
and for M-type stars (to account for the differences in the spectra). In
some cases, in particular for young stars in regions affected by nebular
emission, or when residual spurious features were present, the automatic
identification of the continuum failed: in such cases, we searched the
maximum in a small interval of $\pm 0.5$~\AA{} around a manually fixed
wavelength position at one or both sides of the lithium line. We caution
that the positioning of the continuum at low S/N ($<20$) may not be accurate
because of the strong noise, and the corresponding EWs may be overestimated
or underestimated by up to 20-30~m\AA. These cases are not specifically
flagged in the final catalogue, but they can be identified using the S/N
values provided in the SNR column.

\begin{table}
\centering
\caption{Components used in the spectral fit for FGK stars
\label{tab:lines}}
\begin{tabular}{llccc}
\hline\hline\noalign{\smallskip}
Component& Element& $\lambda$ (\AA)& UVES& Giraffe \\
\noalign{\smallskip}\hline\noalign{\smallskip}
Li line  & \ion{Li}{i}& 6707.761& \checkmark& \checkmark \\
         & \ion{Li}{i}& 6707.912& \checkmark& \checkmark \\
\noalign{\medskip}
Fe blend & \ion{Fe}{i}& 6707.431& \checkmark& \checkmark \\
         & \ion{V}{i} & 6707.518& \checkmark& \checkmark \\
         & \ion{Cr}{i}& 6707.596& \checkmark& \checkmark \\
\noalign{\medskip}   
Comp. 1  & \ion{Si}{i}& 6708.023& \checkmark&            \\
         & \ion{V}{i} & 6708.094& \checkmark&            \\
\noalign{\medskip}
Comp. 2  & \ion{Fe}{i}& 6708.282& \checkmark& \checkmark \\
         & \ion{Fe}{i}& 6708.347& \checkmark& \checkmark \\
\noalign{\smallskip}\hline
\end{tabular}
\end{table}

\begin{figure*}
\centering
\resizebox{\hsize}{!}{\includegraphics{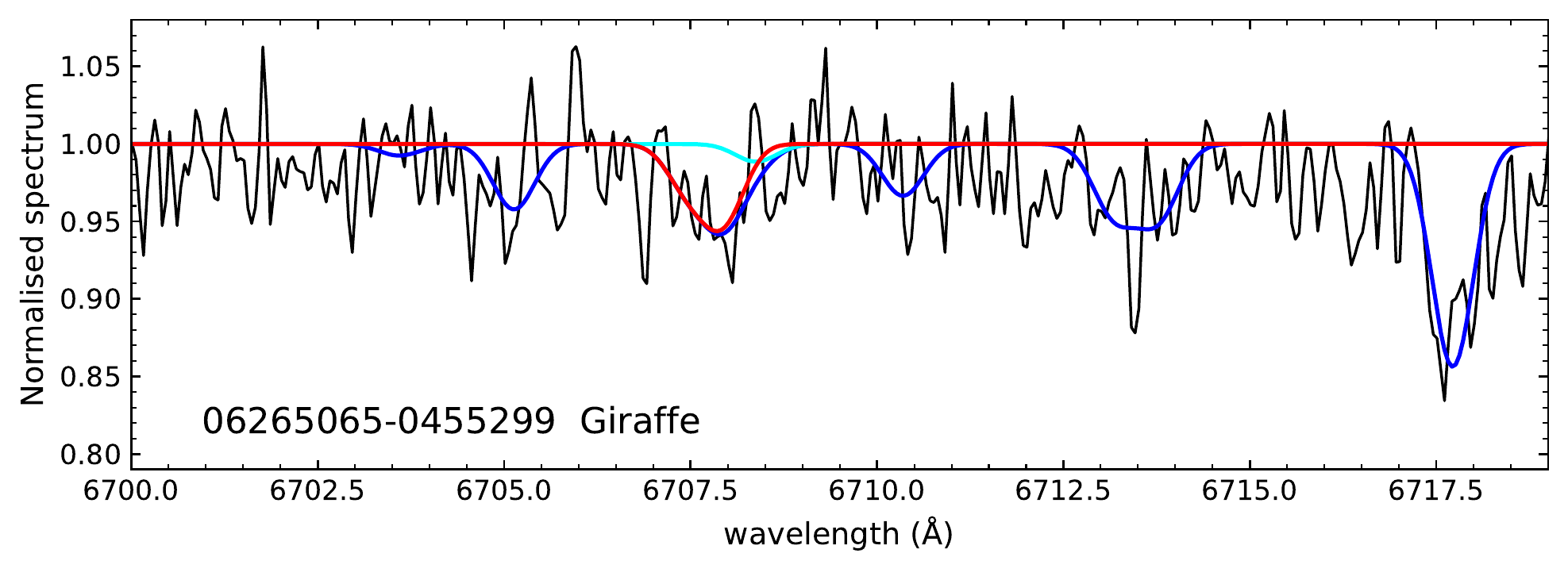}}
\caption{Best-fit result for the spectrum of the metal-poor star
06265065-0455299 observed with Giraffe (S/N\,=\,33). This star has
$T_\mathrm{eff}=5565\pm 95$~K, $\log g = 3.79\pm 0.39$,
$\mathrm{[Fe/H]}=-0.74\pm 0.14$, EW(Li+Fe)$\,=52.4\pm 8.7$~m\AA, and
$A(\mathrm{Li})=2.10\pm 0.13$. Line colours are the same as the bottom panel
of Fig.~\ref{fig:fit_ug}.}
\label{fig:fit_mpoor}
\end{figure*}

\subsection{EW measures}
\label{sec:ews}

In the case of FGK stars, EWs were measured by fitting the spectra,
normalised by the local continuum and shifted to the rest wavelength, with a
combination of Gaussian components. We adopted the same code for UVES and
Giraffe, with only some slight differences due to the different spectral
resolution of the two instruments. The fit was performed using the {\tt
lmfit}\footnote{Available at \url{https://lmfit.github.io/lmfit-py/}} 
python package \citep{lmfit}, which provides a simple framework to perform a
non-linear least-squares fit of complex models with several components,
allowing users to easily change, fix or tie the different model parameters.
This method was applied for $T_\mathrm{eff}\ge 4000$~K to all UVES spectra
and to Giraffe spectra with $\mathrm{[Fe/H]}<-1.5$, for which M-type COGs
were not computed and only FGK COGs are available. For the remaining Giraffe
spectra, we used this method for stars with $T_\mathrm{eff} > 4250$~K, that
is, we set the threshold between the FGK and M regimes in the middle of the
overlapping region of the two COGs. This choice allowed us to ensure better
continuity between the measures in the two regimes, and to derive the
abundances and their uncertainties using the same COG. 

The components used for the fit around the lithium position are consistent
with the linelist of \citet{guiglion16}, which was used to generate the
synthetic spectral grid, and are indicated in Table~\ref{tab:lines}. For
lithium, we fitted the two components of the doublet separately. The Fe
blend includes two additional lines besides the \ion{Fe}{i} line, which are
generally weak but contribute to the global line shape. In addition, we also
included two other components at $\sim$\,6708.0 and $\sim$\,6708.3~\AA.
These components amount generally to a few m\AA, but can become significant
as temperature decreases at solar and super-solar metallicities and/or for
enhanced abundances with respect to solar of the corresponding species, and
they are required to fit correctly the lithium line at low abundances. The
component at $\sim$\,6708.0~\AA{} was only used for UVES, since at the
Giraffe resolution it is completely blended with lithium and cannot be
constrained. In the fit, the widths of all lines were tied together, and we
fixed their relative position with respect to the first line of the lithium
doublet, allowing the latter to vary slightly in wavelength to allow for
possible uncertainties in the radial velocity correction. In addition to the
lines listed in Table~\ref{tab:lines}, a set of relatively strong Fe lines
between 6703.6 and 6713.7~\AA, and, for Giraffe only, the \ion{Ca}{i} line
at 6717.69~\AA, were also included in the fit to better constrain the
Gaussian widths. An example of the fit for UVES and Giraffe is shown in
Fig.~\ref{fig:fit_ug}, where we plot the best-fit results for the spectra of
Fig.~\ref{fig:compare_ug}. In Fig.~\ref{fig:fit_mpoor} we show a more
problematic case, for a metal-poor star observed with Giraffe at
S/N$\sim$30, where the fit is constrained by the \ion{Ca}{i} line.

The EWs of each component were derived from the amplitudes of the best-fit
Gaussians. In the case of UVES, the Li-only EW was simply obtained by
summing the two lithium components. In a few cases, where the lithium line
was very strong, the fit failed, artificially enhancing the 6708.0~\AA{}
line and reducing the Li contribution\footnote{This is due to the fact that
the line amplitudes are free to vary independently from each other.}: 
when this happened, the EW of the 6708.0~\AA{} line was combined with the Li
EW to produce the final measure. In the case of Giraffe, or for UVES spectra
with high rotation, the blended Li+Fe EW was obtained by combining the EWs
of the Li and Fe components; for UVES, we also added the 6708.0~\AA{}
component, which is generally zero but sometimes, when the line is strong,
is enhanced by the fit compensating the lithium line, as mentioned above.
Uncertainties on the measured EWs were computed by means of the
\citet{cayrel88} formula, which however assumes no uncertainty in the
continuum placement, using the FWHM derived from the best-fit. For a few
stars with large rotation rates observed with Giraffe, the Gaussian fit was
not acceptable: in these cases, a direct integration\footnote{For direct
integration, we used the trapezoidal rule implemented in {\tt numpy.trapz}.} 
was used, adopting the same interval defined for M-type stars, and
uncertainties were computed using the error propagation formula on the EW.

In M-type stars, the pEW were measured by direct integration over the
interval defined in Sect.~\ref{sec:cogs_M}. As mentioned above, this method
was used for Giraffe spectra with $T_\mathrm{eff} \le 4250$~K and
$\mathrm{[Fe/H]}\ge-1.5$. We checked that above this temperature threshold
the integrated pEW is consistent, within a few m\AA, with the EW obtained
using the Gaussian fit for spectra with S/N$\,>30$ and low rotation rates.
At lower S/N, noise starts to affect the integrated values, while at high
rotation rates the widening of the integration interval starts to include
additional components that may not be fully taken into account in FGK stars.
Direct integration was also used for stars with no derived $T_\mathrm{eff}$
that showed a clear M-type spectrum. Uncertainties were derived using the
error propagation formula on the EW. 

\begin{figure}
\centering
\resizebox{\hsize}{!}{\includegraphics[clip]{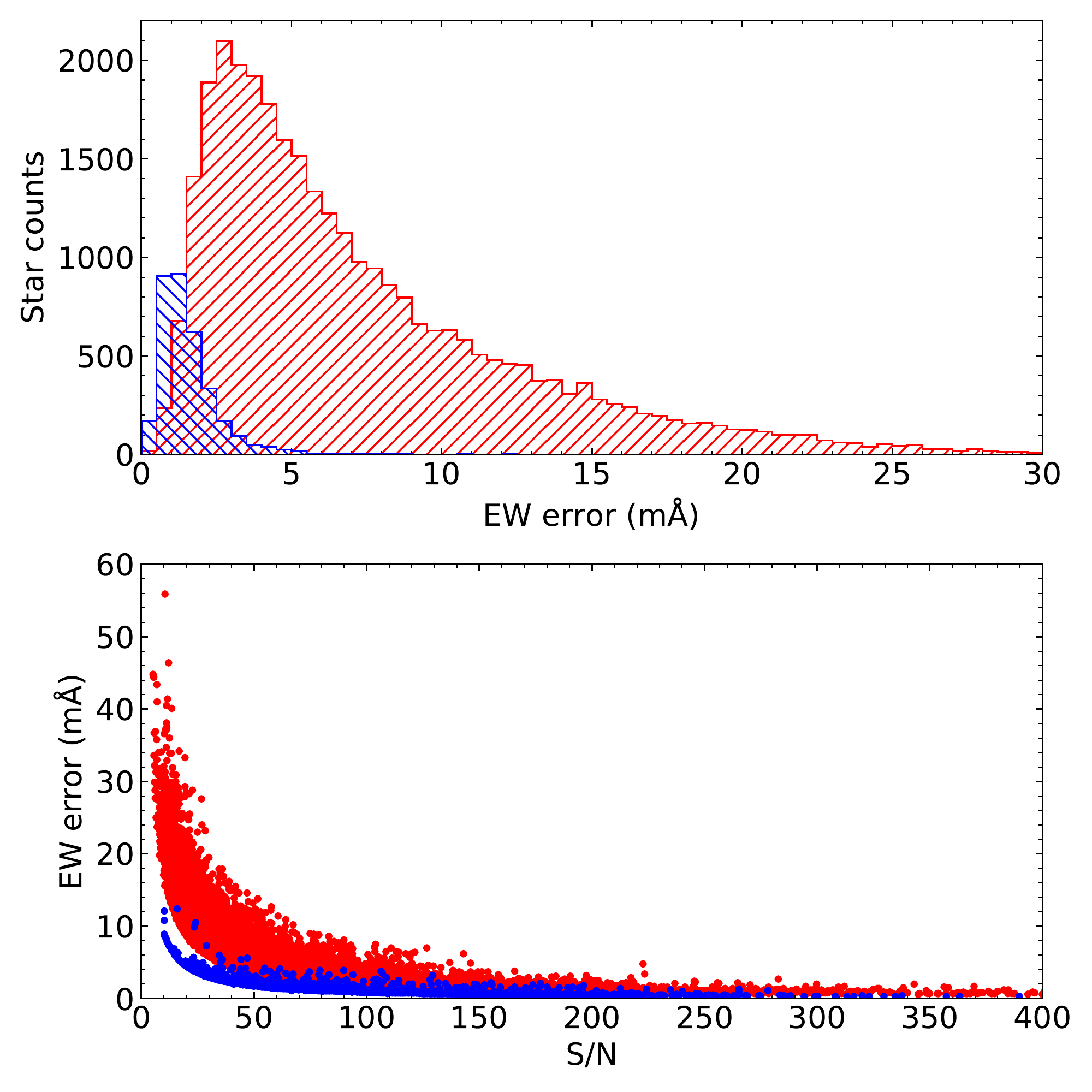}}
\caption{Uncertainties on EW or pEW. In the top panel, we show the
distribution of the uncertainties for Giraffe (red hatched histogram) and
UVES (blue hatched histogram). In the bottom panel, we plot the
uncertainties on EW or pEW as a function of the S/N. Colours are the same as
the top panel, and the plot was limited to S/N$\,<400$ for better clarity.
Uncertainties are higher for stars with higher rotational velocity.}
\label{fig:errors_ew}
\end{figure}

Figure~\ref{fig:errors_ew} shows the distribution of the derived
uncertainties on EW or pEW and their dependence on S/N for Giraffe and UVES
separately. In the case of UVES, uncertainties are generally lower than
5~m\AA{} (except for S/N\,$<20$, where they can reach $\sim$\,10~m\AA), with
the bulk of them concentrated between 1 and 2~m\AA. Uncertainties on Giraffe
EWs are larger, with a median of $\sim$\,6~m\AA{} and a long tail reaching
some tens of m\AA{} for low S/N spectra.

\subsection{Upper limits}
\label{sec:uls}

Upper limits were automatically assigned whenever the measured EW or pEW was
lower than the corresponding uncertainty. In these cases, the uncertainty
value was taken as upper limit. We further visually inspected all cases with
S/N$\,<30$ or EW$\,\la 50$~m\AA{}, and assigned upper limits equal to the
measured EW when the line was not or barely visible.

Since the visual identification of upper limits may be subjective and prone
to errors, especially in noisy spectra, in the case of UVES we devised a
procedure that allowed us to identify secure upper limits and detections in
a semi-automatic way. To this aim, we computed the difference between the
medians of the residuals of the fit obtained with and without the Li
component in the interval [6707.7,\,6707.95]~\AA{}, and we divided it by the
standard deviation of the residuals of the fit with lithium. We assumed this
standard deviation to represent a reasonable estimate of the noise level,
instead of considering a continuum region free of lines, because of the
difficulty to find a suitable region valid for the large variety of GES
spectra. This procedure was tested on a sub-sample of spectra, and we found
that when the derived ratio is $>2$ the lithium line is generally present,
while when the ratio is $<1.7$ for giants and $<1.4$ for dwarfs the line is
generally undetectable. Between these two thresholds most cases are upper
limits, but detections can also be found, and visual inspection was still
required. Applying the above criteria to the whole sample of UVES spectra
allowed us to identify additional upper limits that were previously
considered as dubious detections. This procedure is not entirely free from
errors, and we cannot exclude that some misclassifications might still be
present, however we estimate that less than 1\% of the sample might be
affected.

The extension of the method described above to Giraffe spectra is not
straightforward, because the blend can be detected when the Fe line is
visible even if lithium is not, and it was not possible to find a reasonable
criterion to identify secure upper limits. Therefore, in the case of
Giraffe, upper limits were only assigned by visual inspection, and in
doubtful cases the measures were generally kept as detections. We caution
that, because of this, it is possible that some of the detections with low
EW at S/N$\,<30$ might have been misclassified and be upper limits instead.

\begin{figure}
\centering
\resizebox{\hsize}{!}{\includegraphics[clip]{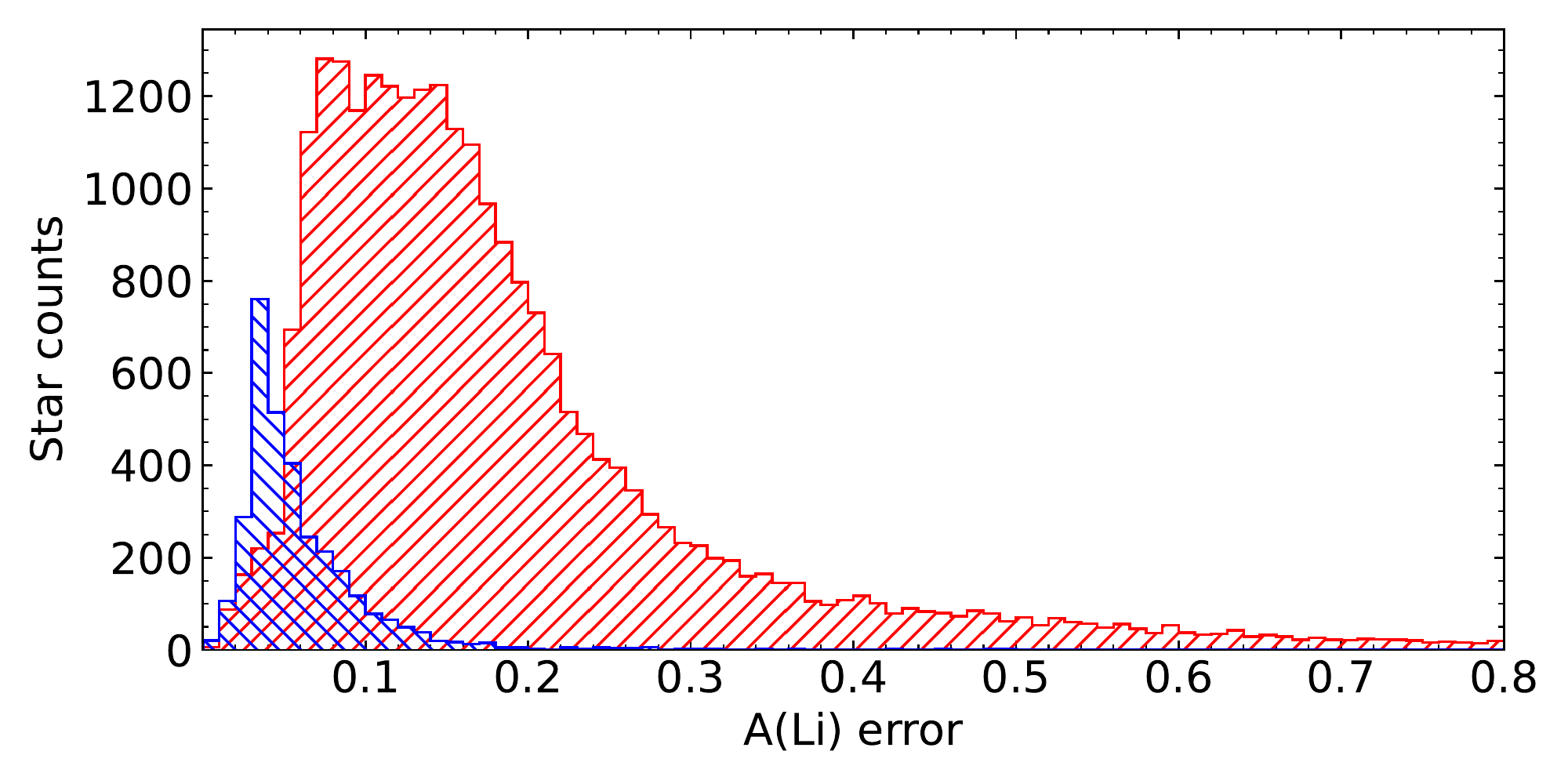}}
\caption{Distribution of the uncertainties on $A$(Li) for Giraffe (red
hatched histogram) and UVES (blue hatched histogram).}
\label{fig:errors_ali}
\end{figure}

\subsection{Abundances}
\label{sec:abund}

Lithium abundances were derived by interpolating the appropriate COGs at the
temperature, gravity, and metallicity (and rotation for M-type stars) of
each star, using the atmospheric parameters derived in the first pass of the
spectral analysis \citep[see][]{gilmore22}. The FGK COGs were used for all
UVES spectra with $T_\mathrm{eff}\ge 4000$~K, for Giraffe spectra with
$T_\mathrm{eff}\ge 4000$~K and $\mathrm{[Fe/H]}<-1.5$, and for Giraffe
spectra with $\mathrm{[Fe/H]}\ge -1.5$ and $T_\mathrm{eff}\ge 4250$~K. For
the remaining Giraffe spectra the M-type COGs were applied. 

Abundances were only derived for stars for which atmospheric parameters
falling inside the relevant COG grid are available. In particular, no
abundances were computed if recommended values of $T_\mathrm{eff}$ or $\log
g$ are not present. However, for stars observed with Giraffe HR15N, an
indication of gravity is given by the $\gamma$ spectral index
\citep{damiani14}: in the $\gamma$-$T_\mathrm{eff}$ diagram, giants occupy a
well defined region clearly distinct from dwarfs. Therefore, for stars with
no $\log g$ but with a measured $\gamma$ index, we assumed $\log g=2.5$ if
$T_\mathrm{eff}<5400$~K and $\gamma > 0.98$ \citep[see e.g.][]{bravi18}, and
$\log g=4.5$ otherwise. This is obviously an approximation, in particular
for PMS stars that would have an intermediate $\log g$, and we caution that
the derived abundances may not be accurate, especially if the true $\log g$
differs significantly from the assumed value\footnote{%
For an M-type PMS star, assuming $\log g=4.0$ instead of 4.5 would increase
the abundance by up to $\sim$\,10\%.}. 
For some stars in young open clusters with $\log g>5.0$, we computed the
abundance using $\log g=5.0$.

When a recommended value of [Fe/H] was not available, we assumed a solar
metallicity. While this assumption is reasonable for many of the open
clusters in our sample, it is clearly incorrect for metal-rich or metal-poor
stars, and in particular for stars in globular clusters. This assumption
should not significantly affect results from UVES spectra, since blends are
believed to weakly contribute to the EW measurement, but it could lead to an
overestimate or underestimate of the Fe-blend correction in Giraffe spectra,
and therefore to an underestimated or overestimated abundance. 

In the case of FGK stars, when the deblended lithium line could be directly
measured, abundances were derived by simply interpolating the FGK COGs. If
only the Li+Fe blend could be measured, we first applied the blend
correction to derive the Li-only EW before computing the abundance; for
these stars, both EW(Li+Fe) and EW(Li) are provided. The correction was not
applied to upper limits: in this case we set an upper limit to EW(Li) equal
to the total Li+Fe upper limit. The same upper limit to EW(Li) was also set
when the corrected EW was lower than the uncertainty, and the derived
abundance was considered an upper limit too. For M-type stars, COGs were
applied directly to the measured pEW. In all cases, whenever the EW or pEW
was lower than the minimum allowed value in the interpolated COG, we set an
upper limit $A(\mathrm{Li})<-1.0$. However, the lithium abundance was
extrapolated above $A(\mathrm{Li}) = 4.0$ if necessary.

The spectra of young stars with significant accretion are affected by
veiling: accretion produces an excess continuum which results in a lower
measured EW. For these stars, the measure of the ratio $r$ between the
excess and photospheric continuum is provided by the OACT node in WG12 for
Giraffe spectra \citep[see][]{lanzafame15}. For stars for which $r$ is
significant (i.e. $r$ greater than its uncertainty) we corrected the
measured Giraffe EW or pEW using the formula EW$_\mathrm{true}=
(1+r)\,$EW$_\mathrm{measured}$; the abundance was then computed from the
corrected EW. In these cases, the uncertainty on the corrected EW was
derived by combining the original EW error and the error on $r$ in
quadrature. However, we did not compute any abundance if $r>1$, since these
high values of $r$ might be inaccurate.

Random uncertainties on the abundances (for detections only) were derived
taking into account the uncertainties on EW or pEW and on the stellar
parameters ($T_\mathrm{eff}$, $\log g$ and [Fe/H]). To this aim, we varied
one parameter at a time by its error and derived the corresponding
uncertainty in the abundance. The individual uncertainties were then
combined in quadrature to derive the final uncertainty. The dominant
contribution to the abundance uncertainties is due to the uncertainties in
EW and $T_\mathrm{eff}$, while the effect of the other parameters is
generally small. In Fig.~\ref{fig:errors_ali} we show the distribution of
abundance uncertainties for UVES and Giraffe separately. As expected, UVES
abundances are more precise, with a median uncertainty of $\sim$\,0.05~dex.
The Giraffe distribution is wider, with a median of $\sim$\,0.15~dex, and an
extended tail with a few errors $>1$~dex, although the bulk of measures have
uncertainties lower than 0.3~dex.

\begin{figure}
\centering
\resizebox{\hsize}{!}{\includegraphics[clip]{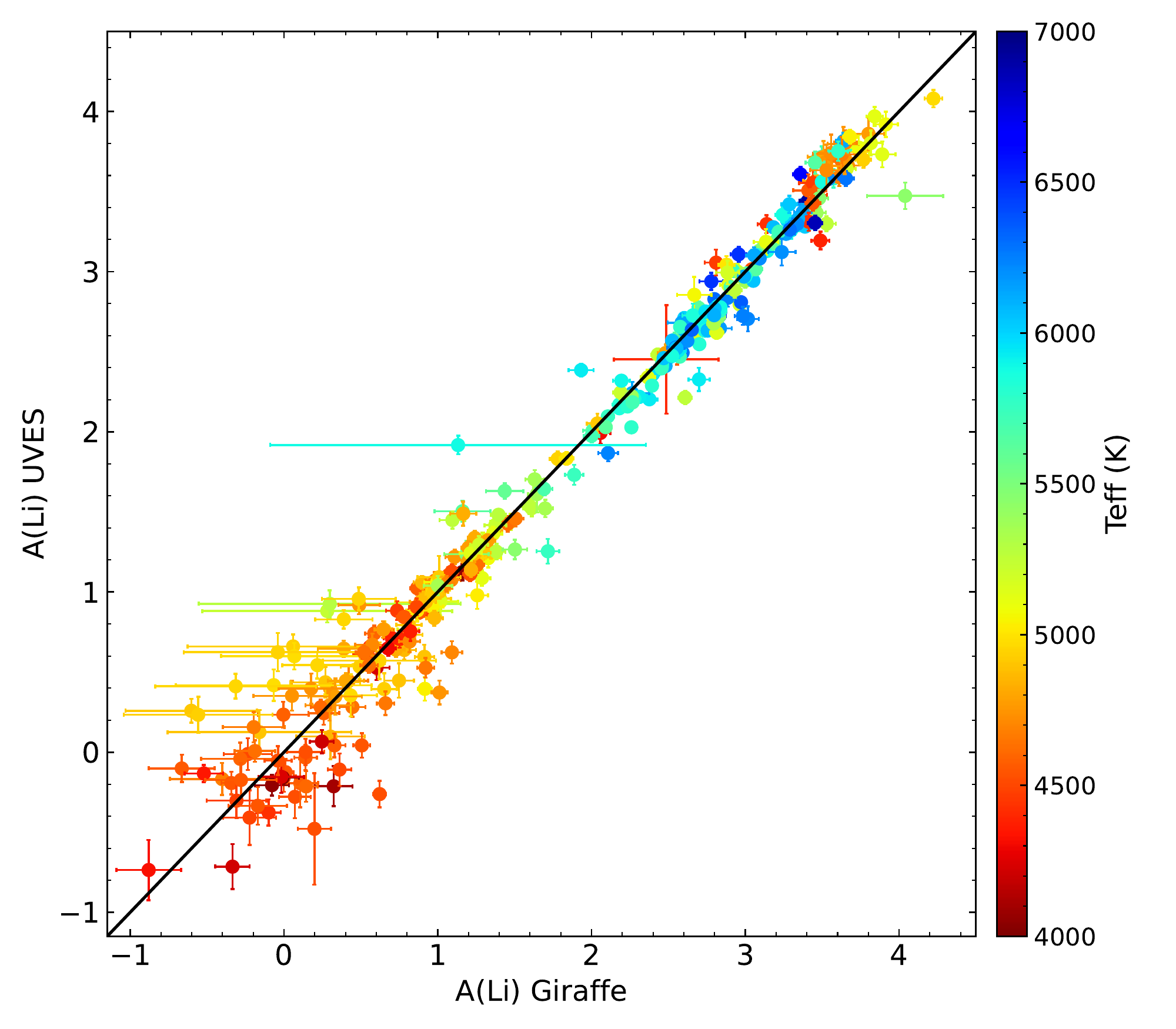}}
\caption{Comparison of the lithium abundances derived for stars observed
with both UVES and Giraffe. Points are colour-coded by $T_\mathrm{eff}$.
Only stars with detections in both instruments are plotted.}
\label{fig:uves-gir}
\end{figure}

In Fig.~\ref{fig:uves-gir} we compare the abundances derived for stars
observed with both UVES and Giraffe, for which lithium was detected in both
instruments. There is in general a very good agreement between the results,
at least for $A(\mathrm{Li})>1.0$, with a larger scatter at lower
abundances. A 3-$\sigma$ clipped average of the differences between UVES and
Giraffe over the entire range gives a mean of $+0.01$~dex with a standard
deviation of 0.13~dex. The standard deviation reduces to 0.10~dex for
$A(\mathrm{Li})>1.0$ and increases to 0.29~dex at lower abundances, with no
significant change in the means ($+0.00$ and $-0.01$~dex, respectively). The
outliers with a large error in Giraffe are mostly giant stars where the
lithium line is very weak and the detection in Giraffe is dominated by the
iron line: in these cases, the measure is likely not accurate, and the
blend-corrected EW is very close to the uncertainty, resulting in large
errors on the derived abundances. 

\begin{figure}
\resizebox{\hsize}{!}{\includegraphics{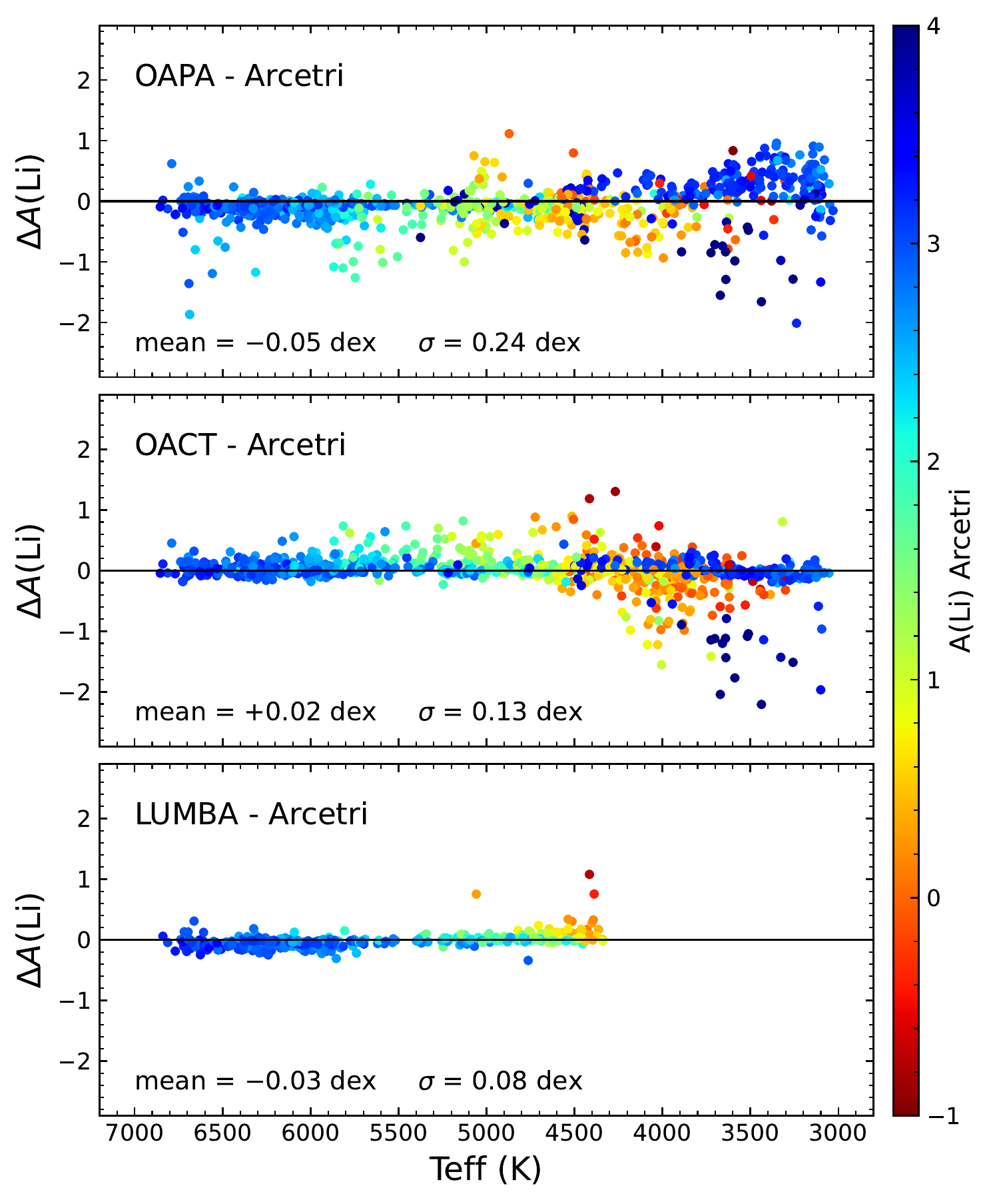}}
\caption{Comparison between the abundances derived by Arcetri with those
derived by other nodes as a function of $T_\mathrm{eff}$, for stars observed
with Giraffe. The plotted differences $\Delta\,A(\mathrm{Li})$ are computed
as the other node minus Arcetri, and are colour-coded by the Arcetri
abundance. Only stars with detections in both nodes are plotted. The
3-$\sigma$ clipped average and standard deviation of the abundance
differences are indicated in the respective panels.}
\label{fig:comp_gir}
\end{figure}
\begin{figure}
\resizebox{\hsize}{!}{\includegraphics{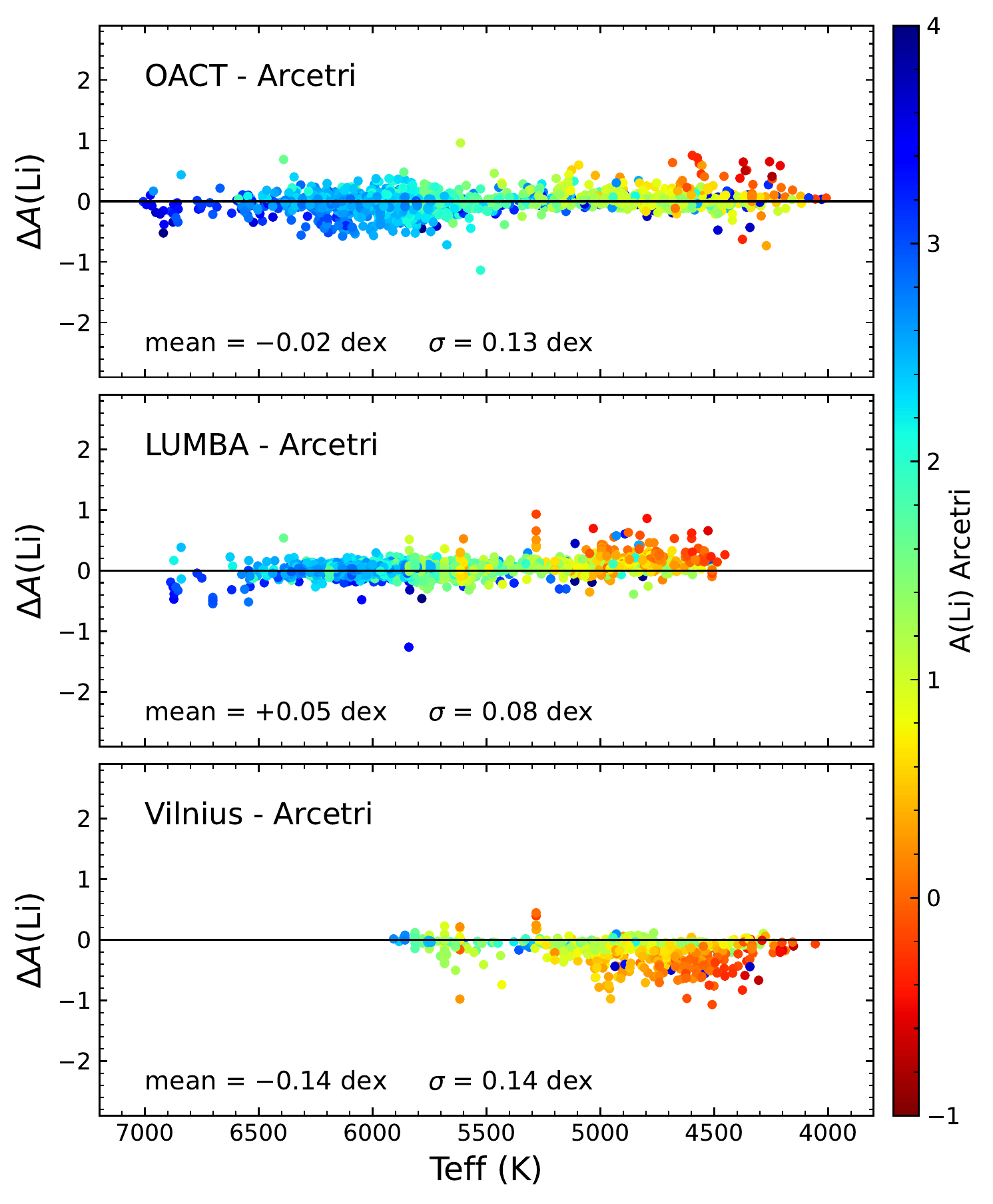}}
\caption{Same as Fig.~\ref{fig:comp_gir} but for stars observed with UVES.}
\label{fig:comp_uves}
\end{figure}


\section{Validation}
\label{sec:validation}

\subsection{Internal validation}
\label{sec:int_val}

To validate our results, we compared them with the abundances provided by
other GES analysis nodes in their respective working groups, namely OAPA
(WG12), OACT (WG10, WG11, and WG12), LUMBA (WG10 and WG11), and Vilnius
(WG11). In the case of UVES, the full dataset was analysed by all
contributing nodes. For Giraffe, the other nodes were asked to provide
measures only for the two open clusters NGC~2420 and NGC~2516 and the PMS
$\lambda$~Ori cluster. The OAPA and OACT nodes measured lithium using the EW
method, with their own codes but following the rules described here and
converting the EWs to abundances using the set of COGs provided in this
paper. The LUMBA and Vilnius nodes used instead spectral synthesis to derive
directly the lithium abundances.

Figure~\ref{fig:comp_gir} shows the differences between the abundances
derived by Arcetri and by the other nodes for Giraffe spectra. There is a
general good agreement between the nodes, although with a larger spread for
the nodes using the EW method. The strong outliers in the comparison with
the OAPA and OACT nodes are generally due to stars with high rotation, low
$T_\mathrm{eff}$, or low S/N, where the measure can be complicated and is
strongly affected by the choice of the continuum, or, in a few cases, to
stars with wrong radial velocity where the lithium line might have been
misidentified. Some of the discrepant cases at $T_\mathrm{eff}<4000$~K are
due to stars affected by veiling that were probably not corrected by the
other nodes. There is a tendency for OAPA to overestimate the abundances of
M-type stars with respect to Arcetri, and for OACT to underestimate them.
However, the differences are relatively uniform at all temperatures and the
average difference is very small in both cases: the 3-$\sigma$ clipped
average is $-0.05$~dex for OAPA and $+0.02$~dex for OACT, with standard
deviations of $\sim$\,0.1$-$0.2~dex. In the case of LUMBA we consider only
FGK stars, since the method used by this node is not able to provide
reliable results for M-type stars. The few outliers are giant stars with
$\log g < 3.1$, for which the LUMBA abundances appear to have been
overestimated, probably because of an erroneous identification of the
lithium detection when the close Fe blend is strong. Apart from these
objects, there is a very good agreement between LUMBA and Arcetri: the
average difference is $-0.03$~dex with a standard deviation of $0.08$~dex.

Figure~\ref{fig:comp_uves} shows the same comparison for UVES spectra.
Again, the agreement with OACT and LUMBA is very good, with average
differences of $-0.02$ and $+0.05$~dex, respectively, and standard
deviations of $\sim$\,0.1~dex. On the other hand, the Vilnius node tends to
provide lower abundances for $T_\mathrm{eff} < 5000$~K, which results in an
average difference of $-0.14$~dex. These stars are mostly giants with very
low lithium abundances, whose measure is strongly dependent on the choice of
the continuum level. Considering only stars with $T_\mathrm{eff} > 5000$~K,
the mean difference reduces to $-0.07$~dex with a dispersion of 0.10~dex.

In summary, the agreement of our abundances with those derived from other
nodes is very good for both instruments, with differences of at most $\pm
0.05$~dex, which are consistent or lower than the average abundance
uncertainties. Such differences can be expected considering differences in
the continuum placement and in the measurement method adopted by the
different nodes.

\begin{figure}
\centering
\resizebox{\hsize}{!}{\includegraphics{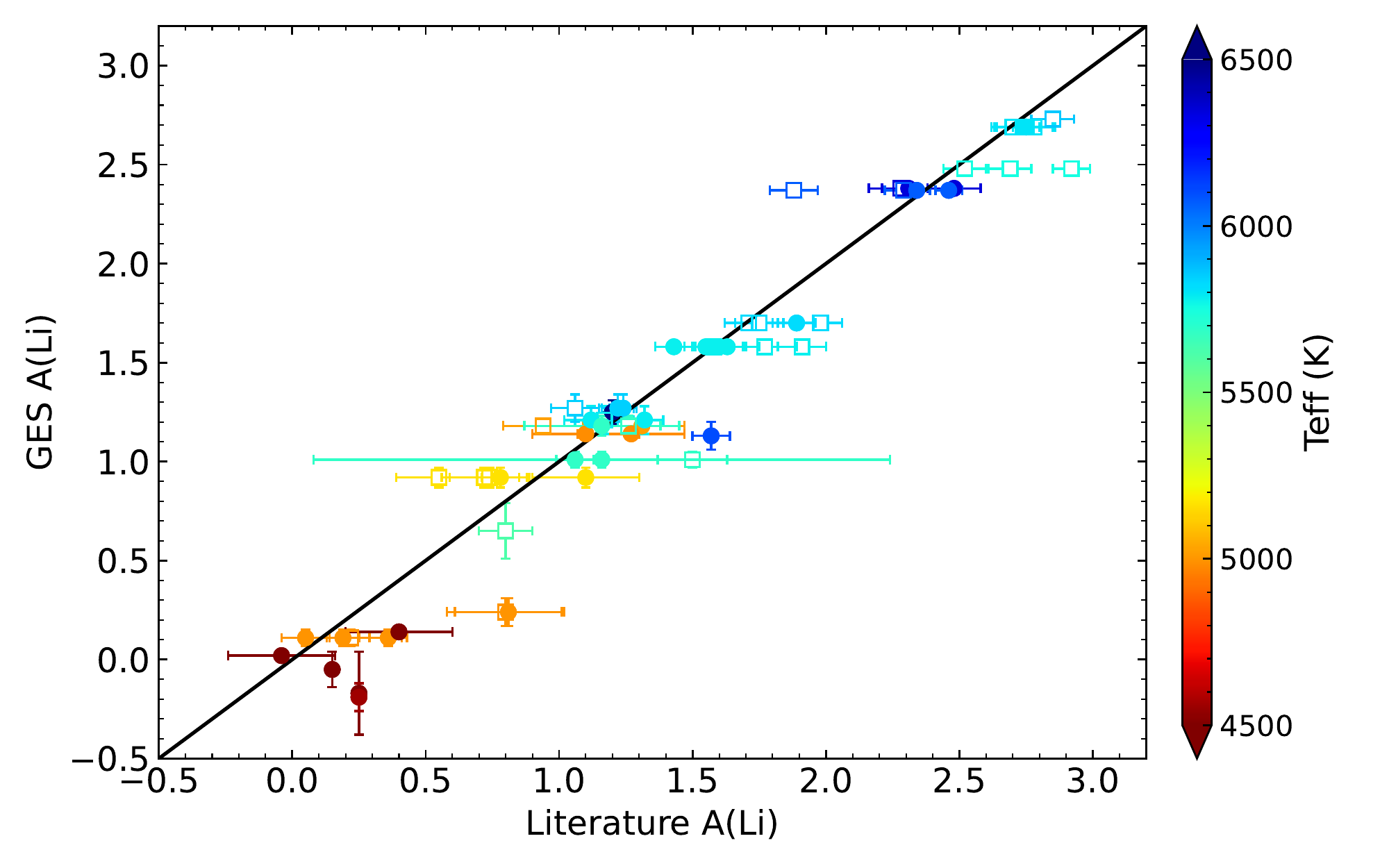}}
\caption{Comparison of the average GES abundances for benchmark stars with
those from the AMBRE project \citep[open squares,][]{guiglion16} and from
other literature studies (filled circles, see text). Only stars with
detections in GES and in the literature studies are plotted. Data are
colour-coded by the GES temperature. Points aligned on the same horizontal
line refer to multiple measures of the same star.}
\label{fig:benchmarks}
\end{figure}

\begin{figure}
\centering
\resizebox{\hsize}{!}{\includegraphics{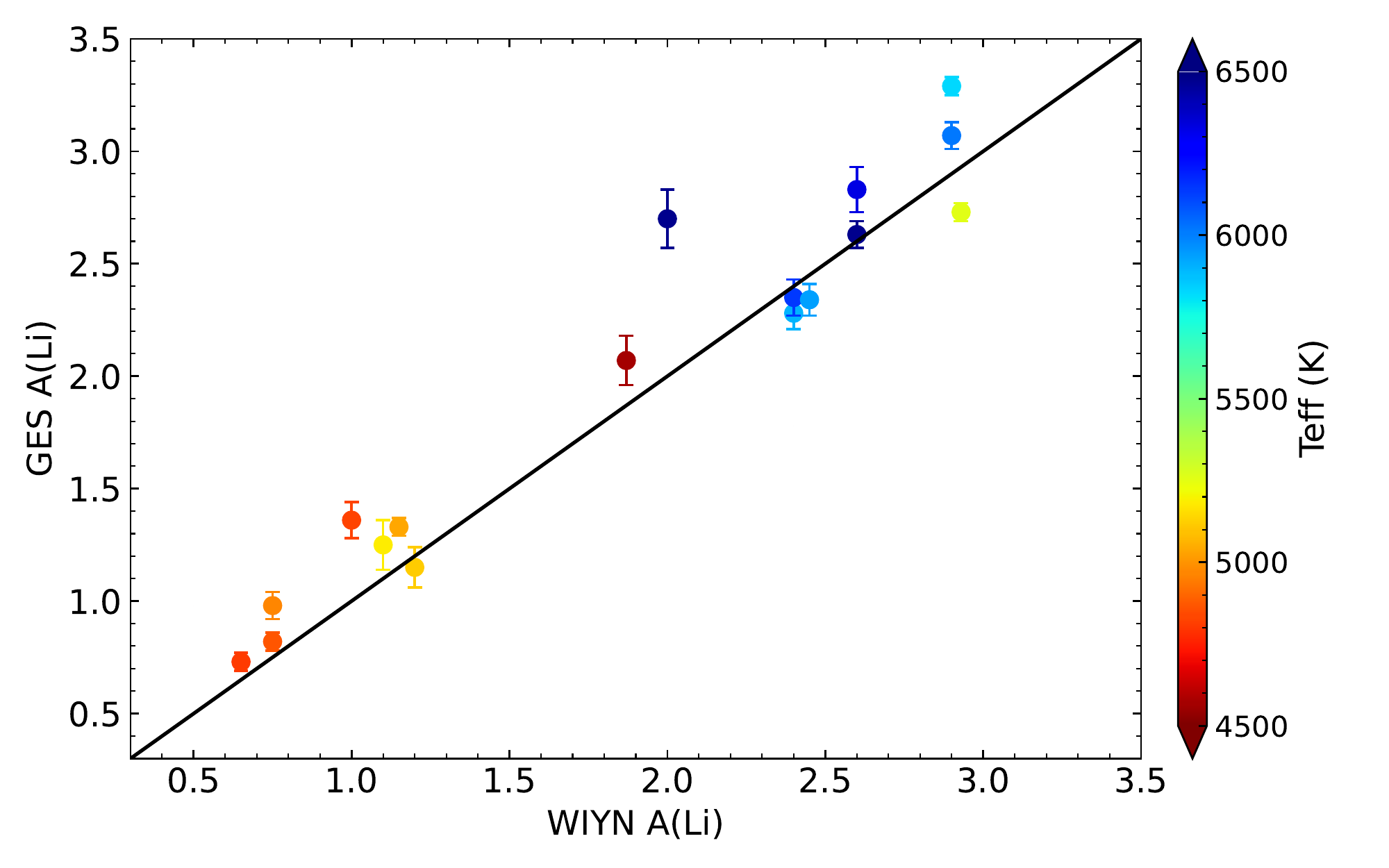}}
\caption{Comparison of the GES recommended abundances with those derived
from the WIYN Open Cluster Study for stars in IC~2391 and NGC~2243
\citep{platais07,at21}. Only stars with detections in both samples are
shown. Data are colour-coded by the GES temperature. Uncertainties on the
WIYN abundances are not available.}
\label{fig:wiyn}
\end{figure}
        
\begin{table*}
\centering
\caption{Lithium-related columns in the final catalogue}
\label{tab:li_columns}
\begin{tabular}{lll}
\hline\hline\noalign{\smallskip}
Column name& Units& Description\\
\hline\noalign{\smallskip}
LI1         & dex& Neutral lithium abundance \\
LIM\_LI1    &    & Flag on LI1 measurement (0=detection, 1=upper limit) \\
E\_LI1      & dex& Error on LI1\\
EW\_LI\tablefootmark{a}& \AA& Total Li(6708\AA) equivalent width or
  pseudo-equivalent width including blends\\
LIM\_EW\_LI &    & Flag on EW\_LI (0=detection, 1=upper limit)\\
E\_EW\_LI   & \AA& Error on EW\_LI\\
EWC\_LI\tablefootmark{b}& \AA& Blends-corrected Li(6708\AA) equivalent width\\
LIM\_EWC\_LI&    & Flag on EWC\_LI (0=detection, 1=upper limit)\\
E\_EWC\_LI  & \AA& Error on EWC\_LI\\
\hline
\end{tabular}
\tablefoot{%
\tablefoottext{a}{Only available for measures from Giraffe or high-rotation
UVES spectra.}
\tablefoottext{b}{Directly measured for UVES, otherwise computed from EW\_LI
using the blend corrections after the eventual correction for veiling. }
}
\end{table*}

\begin{table*}
\centering
\caption{Lithium-specific TECH flags in the final catalogue}
\label{tab:li_flags}   
\begin{tabular}{ll}
\hline\hline\noalign{\smallskip}
Flag\tablefootmark{a}& Description\\
\hline\noalign{\smallskip}
10110-wg-01-01 & Over- or under-subtracted sky features at the position of
   the Li line \\
12003-wg-01-01\tablefootmark{b} & Li abundances are not provided because
   $\log g$ is not available\\
12004-wg-01-00|17103-wg-01-01& Metallicity is not provided and the solar
   value is used\\
12005-wg-01-01& Li not measured because $v\sin i > 50$~km~s$^{-1}$\\
12005-wg-01-02\tablefootmark{c}& Li not measured because $v\sin i >
   100$~km~s$^{-1}$\\
12009-wg-01-01& Li line not measurable in stars with $T_\mathrm{eff} >
   8000$~K\\
12010-wg-01-01& Li abundance is not provided because some parameters are
   outside the COG grid\\
12012-wg-01-01|17103-wg-01-02\tablefootmark{c}& Gravity is not provided and
   it was estimated using the $\gamma$ index (see text)\\
\hline 
\end{tabular}
\tablefoot{
\tablefoottext{a}{wg = 10, 11 or 12 depending on the working group from
which the recommended values were taken.}
\tablefoottext{b}{WG11 only.}
\tablefoottext{c}{WG10 and WG12 only.}
}
\end{table*}

\subsection{External validation}
\label{sec:ext_val}

In addition to the above analysis, we also compared our results for
benchmark stars with the abundances available in the literature. The GES
sample contains a set of UVES and Giraffe spectra of the Sun. We derived an
average lithium abundance of $1.07\pm 0.02$~dex for UVES and $1.06\pm
0.10$~dex for Giraffe. Both values are in very good agreement with the solar
abundance of $1.05\pm 0.10$~dex derived by \citet{grevesse07}. In
Fig.~\ref{fig:benchmarks} we compare the GES recommended abundances (see
Sect.~\ref{sec:catalogue}) of other benchmark stars with those obtained by
the AMBRE project \citep{guiglion16} and by other studies available in the
literature \citep{lebre99,mallik99,takeda07,mentuch08,baumann10,gonzalez10,%
ramirez12,delgado14,delgado15,bl18,chavero19,charbonnel20}. For most stars,
multiple measures are available in the literature, with a significant
spread. However, on average there is an excellent agreement between our
abundances and the literature ones; a 3$\sigma$-clipped average of the
differences between GES and literature values gives a mean of $-0.07$~dex
with a standard deviation of 0.20~dex.

Our sample contains 213 stars with detected lithium abundance that are in
common with the GALAH (GALactic Archaeology with HERMES) DR3 catalogue
\citep{buder21}, considering only GALAH objects with flag\_sp=0 and
flag\_li\_fe=0. However, a direct comparison between the two datasets is not
straightforward, because the abundances provided by GALAH were derived in
non-LTE: to do so, we would have to convert our abundances using non-LTE
corrections, which are model-dependent and could introduce a bias.
Therefore, we decided not to perform the comparison here.

Finally, in Fig.~\ref{fig:wiyn} we compare the GES recommended abundances of
stars in the open clusters IC~2391 and NGC~2243 with those derived in the
context of the WIYN Open Cluster Study \citep{platais07,at21}, for the 17
stars in common with detections in both samples. Also in this case the
agreement is good, although the GES abundances tend to be slightly higher on
average, with a mean difference of 0.13~dex and a dispersion of 0.22~dex.

\begin{table*}
\centering
\caption{Number of lithium detections (det.) and upper limits (u.l.), for
the full sample and divided by target types.}
\label{tab:nmeasures}
\begin{tabular}{llrrrrrrrr}
\hline\hline\noalign{\smallskip}
 & GES\_TYPE\tablefootmark{a}& $N_\mathrm{abund}$\tablefootmark{b}& 
 $N_\mathrm{EW}$\tablefootmark{c} \quad&\multicolumn{2}{c}{LI1}& 
\multicolumn{2}{c}{EW\_LI}& \multicolumn{2}{c}{EWC\_LI}\\
 & & & & det.& u.l.& det.& u.l.& det.& u.l.\\
\hline\noalign{\smallskip}
Total                  &  All             & 38081& 40079& 27250& 10831& 30445& 4002& 23894& 9742\\
\noalign{\medskip}
Open clusters          & *\_CL/*\_SD\_OC  & 30230& 32149& 22639&  7591& 26781& 3873& 19479& 6662\\
Globular clusters      & *\_SD\_GC        &  1240&  1262&   869&   371&   892&   19&   879&  346\\
Milky Way fields       & GE\_MW/GE\_MW\_BL&  3345&  3393&  1521&  1824&     7&    0&  1555& 3087\\
Corot \& Kepler2 fields& GE\_SD\_CR/GE\_SD\_K2& 3206& 3212& 2169& 1037&  2751&  110&  1933&  893\\
Benchmark stars        & others           &    60&    63&    52&     8&    14&    0&    48&    3\\
\hline
\end{tabular}
\tablefoot{
\tablefoottext{a}{* = GE or AR for targets observed by GES or taken from the
ESO archive, respectively.}
\tablefoottext{b}{Total number of stars with derived lithium abundance or
upper limit.}
\tablefoottext{c}{Total number of stars with measured EW and pEW or upper
limit.}
}
\end{table*}

\begin{figure*}
\resizebox{\hsize}{!}{\includegraphics[clip]{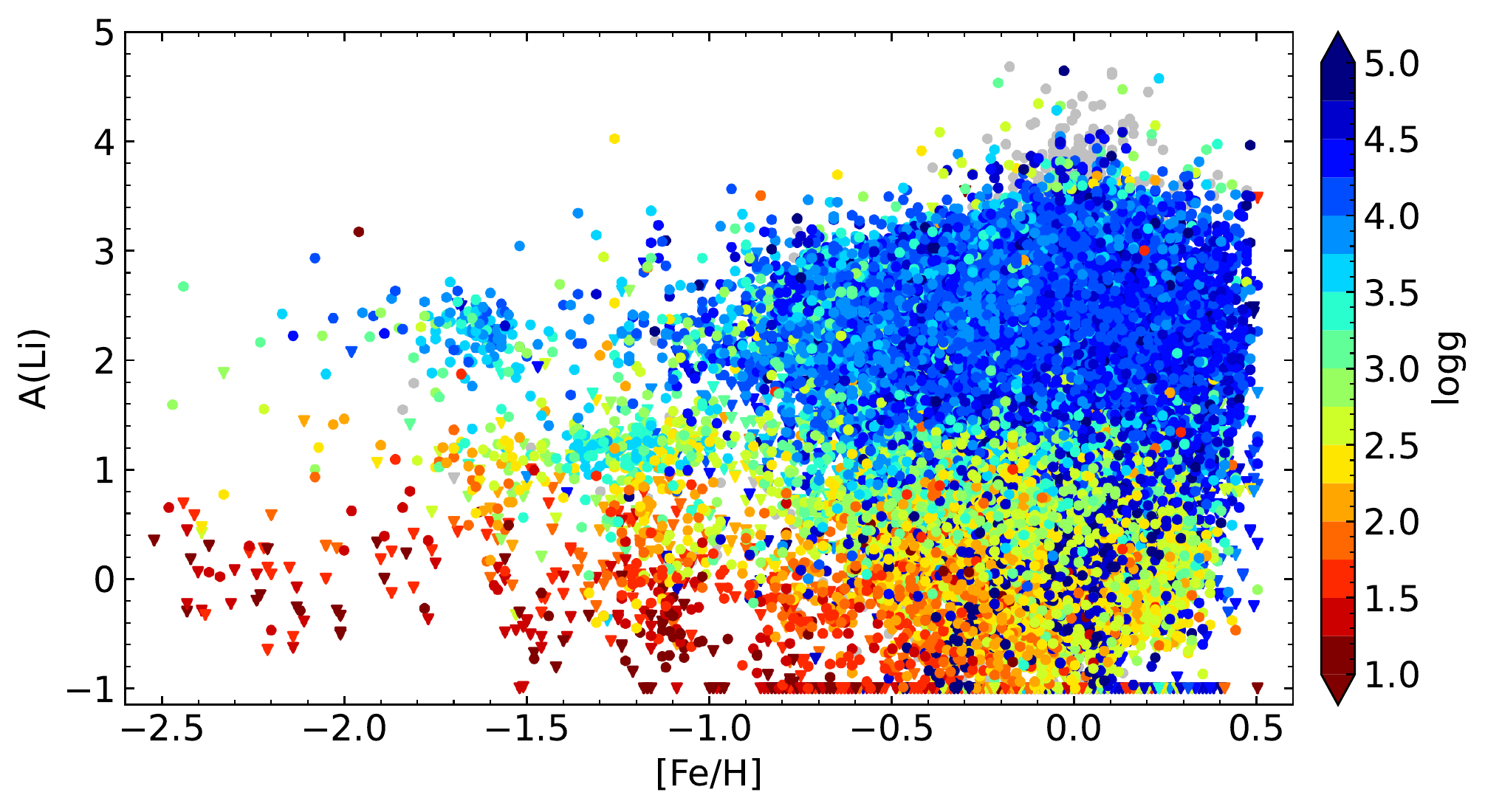}}
\caption{Full sample of lithium abundances as a function of metallicity.
Values are colour-coded by $\log g$. Grey points mark young stars for which
no $\log g$ is available. Downward triangles indicate upper limits.}
\label{fig:allabunds}
\end{figure*}


\section{The final catalogue}
\label{sec:catalogue}

The results obtained in Sect.~\ref{sec:measures} were combined together to
produce the final catalogue. As a first step, for each setup we first
combined the measures obtained from multiple spectra of the same star, which
were available for benchmark stars and for a few open cluster stars with
archival observations in addition to the GES ones. For these cases, if one
or more detections were available, we took as final value the average of the
detections (for both EWs and abundances), and the average error as
uncertainty. This procedure was also applied when a mix of detections and
upper limits for the same star was present. If only upper limits were
present, we took the highest upper limit as final value, to be conservative.
We then merged the results obtained from UVES and Giraffe. UVES measures
were retained only for $T_\mathrm{eff}\ge 4000$~K, since EWs measured at
lower temperatures are not reliable and abundances were not derived. When
results from both instruments were available, we took preferentially the
UVES measures as recommended values, since they are more precise. Otherwise,
the only available UVES or Giraffe measures were taken as recommended
values. 

Table~\ref{tab:li_columns} lists the lithium-related columns available in
the catalogue. Lithium abundances are given in the LI1 column, with
associated columns for the errors and flags. In addition, we also provide
the measured EWs. EW\_LI contains the measured total EW (including blends)
for FGK stars, or the measured pEW for M-type stars with $T_\mathrm{eff}\le
4250$~K and [Fe/H]~$\ge -1.5$. EWC\_LI contains the Li-only EW, either
directly measured in UVES (in this case EW\_LI is empty), or derived from
EW\_LI using the blend corrections after the eventual correction for
veiling. EWC\_LI is empty for M-type stars for which the pEW is given in
EW\_LI, or when the abundance could not be derived. In addition to these
columns, veiling measures are also available in the columns VEIL and
E\_VEIL. The catalogue also contains a set of technical flags in the TECH
column, and peculiarity flags (e.g. for binaries) in the PECULI column. In
Table~\ref{tab:li_flags} we only list the TECH flags specific to the lithium
measurements, but other generic flags on S/N or problems in the spectra may
also be present.

The final catalogue at ESO contains lithium EW measures or upper limits for
a total of 40\,079 stars, and lithium abundances for 38\,081 stars. The
number of available measures for the different sub-sample types is given in
Table~\ref{tab:nmeasures}, and the distribution of lithium abundances as a
function of metallicity and gravity is shown in Fig.~\ref{fig:allabunds}.
The vast majority of the sample ($\sim\,$80\%) consists of stars observed in
the fields of open clusters spanning a large age range, from a few Myr up to
$\sim\,$9~Gyr, and covering a wide range of Galactocentric distances
\citep[see][]{randich22}; in addition, about 3\% of the lithium measures
were obtained for stars in globular clusters. The remaining 17\% are field
stars, including stars in the Bulge and in Corot and Kepler2 fields. The
sample covers well all evolutionary phases in the HR diagram, from PMS stars
to giants, and also includes metal-poor stars on the Spite plateau.
Therefore, this catalogue constitutes an invaluable dataset for the
investigation of various topics, including membership of young clusters,
determination of stellar ages, constraints on models of stellar structure
and evolution, and Galactic evolution.


\section{Caveats}
\label{sec:caveats}

As already mentioned in the previous sections, there are a few caveats that
users of the catalogue should keep in mind when using the GES lithium
dataset. We summarise them below.

The reported EWs for stars observed with Giraffe with $T_\mathrm{eff}\le
4250$~K and [Fe/H]~$\ge -1.5$ are pEWs that include molecular and other line
blends. For this reason, pEWs do not go down to zero and may be significant
even when no lithium is present (see Sect.~\ref{sec:cogs_M}). In addition to
that, lithium-only EWs measured in UVES for $T_\mathrm{eff} \la 4250$~K and
$\mathrm{[Fe/H]}\ga 0$ are strongly affected by blends and may not be
accurate.

As noted in Sect.~\ref{sec:uls}, EW upper limits for Giraffe and part of
those for UVES were assigned after visual inspection of the fitted spectra.
Because of this, some of the measures reported as detection for spectra with
low S/N ($<30$) and weak lithium line (EW$\,\la 50$~m\AA) might have been
misclassified and be upper limits instead.

The cut in $v\sin i$ described in Sect.~\ref{sec:measures} was applied using
the values derived from the data reduction pipeline. Since some rotational
velocities were revised after the spectral analysis, especially for the
hotter stars, the final catalogue contains a few lithium measures with low
EW at $v\sin i > 150$~km~s$^{-1}$. For part of them, the original rotational
velocity was significantly underestimated, and the corresponding lithium
measures might be inaccurate.

As discussed in Sect.~\ref{sec:abund}, if a recommended metallicity was not
available, the solar metallicity was used when computing lithium abundances.
If the true metallicity is significantly different, the use of the solar
value results in wrong abundances. This is especially true in the case of
Giraffe, where the blend correction can be severely over- or underestimated.
A similar problem holds for stars with no recommended $\log g$ value, where
an approximated gravity was assumed based on the $\gamma$ index, if
available: as above, if the true gravity is significantly different from the
assumed one, the corresponding abundance is likely wrong. These cases are
appropriately flagged in the final catalogue (see Table~\ref{tab:li_flags}),
and abundances for these stars should be taken with caution.

The measured Giraffe EWs for young stars affected by accretion were
corrected for veiling before computing the abundances. However, estimating
the veiling is not simple, and some of the reported values may be
inaccurate: in this case, the derived lithium abundance may also be
inaccurate. In particular, if the reported veiling is overestimated, the
abundance will be overestimated as well. For this reason, we did not compute
abundances if the veiling value was $>1$, and abundances for stars with
veiling should be treated with caution.

A few cool stars in young clusters may have an underestimated value of $\log
g$, indicative of giants, in contrast with the derived $\gamma$ index which
is characteristic of dwarfs. Since abundances were derived using the $\log
g$ value, the corresponding abundances for these stars might not be
accurate. An example of this issue are the two members of the 25~Ori cluster
mentioned in Sect.~3.5 of \citet{francio22}, whose abundances might have
been overestimated by $\sim 0.4-0.6$~dex.

Although clear SB2s were discarded from the sample before performing the
measurements (see Sect.~\ref{sec:measures}), some SB2s can only be detected
by a careful exam of the cross-correlation function \citep{merle17}. In
addition, a few SB2s were flagged by other nodes, or might have been
identified in UVES but not in Giraffe for stars observed with both
instruments. For such stars, the PECULI binarity flag 20020 is raised, and
the corresponding lithium measurements, if present, are likely inaccurate.

We finally caution that, if the comparison with evolutionary models is made
using EWs rather than abundances, the COGs provided here should be used to
convert the model abundances into EWs. This is particularly true in the M
dwarf regime; using a different set of COGs would result in inconsistencies
between the measures and the models, leading to possibly inaccurate
conclusions.


\section{Summary}
\label{sec:summary}

This paper describes the derivation of lithium abundances for the final data
release of the GES. Lithium was measured on spectra obtained with both the
Giraffe and UVES instruments, and covering a wide range of temperatures,
gravity, and metallicity. We used an EW-based method, that is, we first
measured the EW of the lithium line, and then we converted it to an
abundance using a set of COGs that were specifically derived for GES. The
COGs were measured on a grid of synthetic spectra covering the full range of
parameters of GES observations, using a method consistent with that adopted
to measure the EWs. The derived abundances are one-dimensional LTE
abundances. We stress that our COGs represent the first set of homogeneously
derived COGs over a wide range of temperatures (3000$-$8000~K), gravities
($\log g=$\,0.5$-$3.5), and metallicities ($-2.50\le \mathrm{[Fe/H]}\le
+0.50$).

The presence of molecular blends in M-type stars, which increasingly affect
the measure of lithium as temperature decreases, forced us to adopt two
different methods for FGK and M-type stars. For FGK stars the lithium EW
could be measured by fitting the line with Gaussian components, while for
M-type stars a pEW was obtained by integrating the spectrum on a predefined
interval. Care was taken to ensure that no significant discontinuity arises
between the two temperature regimes. The derived abundances were validated
using measures provided by other GES analysis nodes or available from the
literature.

The final catalogue includes homogeneous lithium abundances and/or EWs for
$\sim$\,40\,000 stars distributed in all Milky Way components (open and
globular clusters, disc, bulge and halo) and covering all evolutionary
phases, from PMS stars to giants. This dataset will be very valuable for our
understanding of several open issues, from stellar evolution and internal
mixing in stars at different evolutionary stages, to the derivation of
stellar ages, and to Galactic evolution. We finally note that the detailed
work presented here will also be very useful for future large scale
spectroscopic surveys, such as WEAVE and the 4MOST high-resolution stellar
surveys, which will be characterised by a similar resolution as HR15N.


\begin{acknowledgements}
Based on data products from observations made with ESO Telescopes at the La
Silla Paranal Observatory under programmes 188.B-3002, 193.B-0936, and
197.B-1074. These data products have been processed by the Cambridge
Astronomy Survey Unit (CASU) at the Institute of Astronomy, University of
Cambridge, and by the FLAMES/UVES reduction team at INAF/Osservatorio
Astrofisico di Arcetri. These data have been obtained from the Gaia-ESO
Survey Data Archive, prepared and hosted by the Wide Field Astronomy Unit,
Institute for Astronomy, University of Edinburgh, which is funded by the UK
Science and Technology Facilities Council. This work was partly supported by
the European Union FP7 programme through ERC grant number 320360 and by the
Leverhulme Trust through grant RPG-2012-541. We acknowledge the support from
INAF and Ministero dell'Istruzione, dell'Universit\`a e della Ricerca (MIUR)
in the form of the grant "Premiale VLT 2012". The results presented here
benefit from discussions held during the Gaia-ESO workshops and conferences
supported by the ESF (European Science Foundation) through the GREAT
Research Network Programme.
We acknowledge the support from INAF in the form of the grant for mainstream
projects ``Enhancing the legacy of the Gaia-ESO Survey for open clusters
science''.
R.S. acknowledges support from the National Science Centre, Poland
(2014/15/B/ST9/03981).
T.B. was supported by grant No. 2018-04857 from the Swedish Research
Council.
M.B. is supported through the Lise Meitner grant from the Max Planck
Society. We acknowledge support by the Collaborative Research centre SFB 881
(projects A5, A10), Heidelberg University, of the Deutsche
Forschungsgemeinschaft (DFG, German Research Foundation). This project has
received funding from the European Research Council (ERC) under the European
Union’s Horizon 2020 research and innovation programme (Grant agreement No.
949173).
This work made use of {\sc Astropy} (http://www.astropy.org), a
community-developed core Python package for Astronomy
\citep{astropy13,astropy18}.
\end{acknowledgements}

\bibliographystyle{aa} 
\bibliography{biblio}


\begin{appendix}
\section{Derived curves of growth}

In Tables~\ref{tab:cog_fgk}-\ref{tab:cog_m} (available in full at the CDS)
we provide the derived COGs and blend corrections for FGK stars, and the
COGs for M-type stars. The first ten lines of each table are shown here for
reference.

\begin{table}[!h]
\centering
\caption{COGs for FGK stars  
\label{tab:cog_fgk} }
\begin{tabular}{cccrr}
\hline\hline\noalign{\smallskip}
$T_\mathrm{eff}$& $\log g$& [Fe/H]& $A(\mathrm{Li})$& EW(Li)\\
(K)&  &  &  & (m\AA)\\
\hline\noalign{\smallskip}
4000& 0.5&  $-$2.50& $-$1.0&   5.38\\ 
4000& 0.5&  $-$2.50& $-$0.8&   8.47\\ 
4000& 0.5&  $-$2.50& $-$0.6&  13.28\\
4000& 0.5&  $-$2.50& $-$0.4&  20.66\\
4000& 0.5&  $-$2.50& $-$0.2&  31.78\\
4000& 0.5&  $-$2.50&    0.0&  48.06\\
4000& 0.5&  $-$2.50&    0.2&  70.95\\
4000& 0.5&  $-$2.50&    0.4& 101.78\\
4000& 0.5&  $-$2.50&    0.6& 139.13\\
4000& 0.5&  $-$2.50&    0.8& 181.39\\
\ldots& \ldots& \ldots& \ldots& \ldots\\
\hline
\end{tabular}
\end{table}

\begin{table}[!h]
\centering
\caption{Blend corrections for FGK stars   
\label{tab:fecorr} }
\begin{tabular}{cccr}
\hline\hline\noalign{\smallskip}
$T_\mathrm{eff}$& $\log g$& [Fe/H]& EW(Fe)\\
(K)&  &  & (m\AA)\\
\hline\noalign{\smallskip}
4000& 0.5&  $-$2.50&   0.95\\
4000& 0.5&  $-$2.00&   2.43\\
4000& 0.5&  $-$1.50&   6.00\\
4000& 0.5&  $-$1.00&  14.56\\
4000& 0.5&  $-$0.75&  23.26\\
4000& 0.5&  $-$0.50&  36.49\\
4000& 0.5&  $-$0.25&  55.68\\
4000& 0.5&  $+$0.00&  82.19\\
4000& 0.5&  $+$0.25& 104.29\\
4000& 0.5&  $+$0.50& 127.48\\
\ldots& \ldots& \ldots& \ldots\\
\hline
\end{tabular}
\end{table}

\begin{table}[!t]
\centering
\caption{COGs for M-type stars 
\label{tab:cog_m}}
\begin{tabular}{ccccrr}
\hline\hline\noalign{\smallskip}
$T_\mathrm{eff}$& $\log g$& [Fe/H]& $v\sin i$& $A(\mathrm{Li})$& pEW\\
(K)&  &  & (km~s$^{-1}$)&  & (m\AA)\\
\hline\noalign{\smallskip}
3000& 0.5&  -1.50& 0.0&  $-$1.0& 469.70\\
3000& 0.5&  -1.50& 0.0&  $-$0.8& 482.25\\
3000& 0.5&  -1.50& 0.0&  $-$0.6& 498.65\\
3000& 0.5&  -1.50& 0.0&  $-$0.4& 518.97\\
3000& 0.5&  -1.50& 0.0&     0.0& 568.76\\
3000& 0.5&  -1.50& 0.0&     0.2& 595.87\\
3000& 0.5&  -1.50& 0.0&     0.4& 622.93\\
3000& 0.5&  -1.50& 0.0&     0.6& 649.38\\
3000& 0.5&  -1.50& 0.0&     0.8& 675.05\\
3000& 0.5&  -1.50& 0.0&     1.0& 700.05\\
\ldots& \ldots& \ldots& \ldots& \ldots& \ldots\\
\hline
\end{tabular}
\end{table}

\end{appendix}

\end{document}